\documentclass[12pt]{article}
\usepackage{epsfig}
\usepackage{wrapfig}
\usepackage{amssymb}
\usepackage{cite}

  \newcommand{\beq}{\begin{equation}}
  \newcommand{\eeq}{\end{equation}}

  \newcommand{\beqa}{\begin{eqnarray}}
  \newcommand{\eeqa}{\end{eqnarray}}

\def\JNA{{J.\ Num.\ Anal.\ }}

\def\NP{{Nucl.\ Phys.\ }}

\def\PL{{Phys.\ Lett.\ }}
\def\PR{{Phys.\ Rev.\ }}

\def\PRL{{Phys.\ Rev.\ Lett.\ }}

\def\lsim{\raise0.3ex\hbox{$<$\kern-0.75em\raise-1.1ex\hbox{$\sim$}}}

\begin{document}

\thispagestyle{empty}

\hfill \begin{minipage}[t]{4cm}
       \begin{flushright}
       BI-TP 2001/03 \\
       \today \\
       \end{flushright}
       \end{minipage}

\vspace*{2cm}

\begin{center}
{\Large\bf
Meson Screening Masses at high Temperature in quenched QCD with
improved Wilson Quarks
}
\end{center}

\vspace*{0.3cm}

\begin{center}
Edwin Laermann and Peter Schmidt \\[0.3cm]
{\it Fakult\"at f\"ur Physik, Universit\"at Bielefeld,
D-33615 Bielefeld, Germany}
\end{center}

\vspace*{3cm}

{\bf Abstract}
  We report on a lattice investigation of 
  improved quenched Wilson fermions above and below the 
  confinement-deconfinement phase transition. Results on
  meson screening masses as well as spatial wave functions
  are presented. Moreover, the meson dispersion relation
  is studied. Below the critical temperature we do not observe
  any significant temperature effect while above $T_c$ the data are
  consistent with a leading free quark behavior.

\newpage

\section{Introduction}

An important goal of analytical as well as lattice investigations
of QCD at non-vanishing temperature has been to gain more
insight into the temperature dependence of hadron properties
below and
into the nature of hadronic excitations above the
transition temperature from the hadronic to the 
plasma phase of QCD.

When the temperature is raised towards the transition point,
approaching (approximate) chiral symmetry restoration
and deconfinement is expected to change
the properties of hadrons. 
In particular, the lightest vector mesons and
the temperature dependence of their
masses and decay widths 
have received quite some attention because of
their possible relevance to the observed enhancement
of low mass dileptons in heavy ion collisions
\cite{dilepton_exp}.
The theoretical predictions of these properties are,
however, model dependent, see e.g. \cite{Rapp} for
a recent review.
In the plasma phase, the effective, temperature dependent coupling 
constant becomes small at large temperatures. One is thus lead to 
expect that the plasma consists of a gas of only weakly interacting 
quarks and gluons. 
In this case, correlation functions of operators with hadron quantum
numbers should be described by the exchange of two or three
almost free quarks.
On the other hand, there are arguments that 
even at high temperature the hadronic excitation spectrum might be more
complicated because of non-perturbative effects in particular in
the chromo-magnetic sector of QCD \cite{DeTar}.
Thus, in both temperature regimes ab-initio QCD computations are
highly desirable.

Hadronic correlation functions at non-vanishing temperature $T$ have
been the subject of lattice investigations for quite a while.
Most of these studies are based on the staggered quark formulation
\cite{Detar2,gocksch,Gottlieb,MTc,Sourendu,Boyd3,Bernard2,Boyd2,Boyd1,
Gottlieb2,Lagae}
while only few so far utilized the Wilson discretization
\cite{Nucu,Nucu2,QCDTARO} and, just recently, domain wall
fermions \cite{Columbia}.
Moreover, because of the limited extent of the lattice in the
Euclidean time direction, $0 < t < 1/T$, most of these studies 
investigated spatial correlation functions and screening masses.

Below the transition temperature,
detailed lattice investigations and comparisons of hadron masses
at $T=0$ and at $T<T_c$ have been carried out in the
quenched approximation so far. 
Analyses in the staggered fermion discretization 
\cite{Sourendu,Boyd1} have covered temperatures
between $0.8 T_c$ and $0.95 T_c$ and did not detect any
significant difference between zero-temperature
masses and finite-temperature screening masses in the investigated 
hadron channels.
This is also supported by a recent study of spatial and temporal
correlation functions of Wilson fermions on anisotropic
lattices \cite{QCDTARO} at $T \simeq 0.93 T_c$.

At temperatures above $T_c$, the available lattice results
on hadronic correlation functions 
reflect chiral symmetry restoration because
the masses \cite{QCDTARO} and screening masses
\cite{Detar2,gocksch,Gottlieb,MTc,Sourendu,Bernard2,
Gottlieb2,Lagae} 
in the vector ($\rho$) and axial vector ($a_1$) channel
become degenerate. In the pseudoscalar and
scalar channels it is observed that $\pi$ and $f_0/\sigma$
become degenerate at $T_c$ \cite{Lagae} while the mass
difference between $\pi$ and $a_0/\delta$ seems to vanish
only at higher temperatures, see in particular \cite{Columbia}
for a chiral extrapolation at temperatures closely above $T_c$.
These findings are in accord with the expected restoration of 
$SU_R(N_F) \times SU_L(N_F)$ and indicate that the anomalous $U_A(1)$
symmetry is not effectively restored at the critical temperature.

As far as the values are concerned,
in the staggered discretization
vector and axial vector screening masses are 
compatible with the prediction of
lowest order perturbation theory i.e. 
the propagation of (almost) free quarks. Scalar and
pseudoscalar channels, however, 
show substantial deviations from
this expectation, at least in the temperature interval between
$T_c$ and $2 T_c$. One might argue that this observation
indicates the existence of (pseudo) scalar
bound states.
However, the lowest mass hadron is unlikely
to change from a mesonic state to a quark-like quasi-particle at
a non-critical temperature. In addition, studies which apply different
boundary conditions \cite{Boyd2} suggest that the spectral
function is dominated by a two quark cut. 
For Wilson quarks it is observed \cite{QCDTARO} that
already at a temperature of about $1.5 T_c$ the 
pseudoscalar and vector screening masses 
are very close to each other.
There it also is found
that a near degeneracy holds true for the masses. Depending on the
source operator utilized, the pseudoscalar is sometimes even
heavier than the vector meson.
While this behavior can
at least qualitatively be explained by 
the propagation of (almost) free quarks, 
\cite{QCDTARO}
provides conflicting evidence since their study of the
wave functions, on the other hand, indicates the presence of bound states.

In addition to these observables, also spatial wave functions
have been analyzed \cite{MILC_BS}.
Here one has found a similar behavior as at zero temperature.
The observed exponential decay has then been taken as to
suggest that the relevant hadronic excitations
are bound states also in the plasma phase, at least
at temperatures just above $T_c$. According to \cite{Koch},
this behavior could, however, also be explained by the fact that
the dimensionally reduced, 3-D effective theory and
correspondingly spatial
Wilson loops in 3+1 dimensions show confinement 
\cite{Borgs,Janos,Martin}.
Solving
a two-dimensional Schr\"odinger equation with a potential
which includes a temperature dependent (spatial) string tension
leads to exponentially decreasing spatial wavefunctions.
The corresponding effect on the screening masses \cite{Hansson}
would be an ${\cal O}(10 \%)$ correction 
at the investigated temperatures \cite{Martin}
which so far could not yet be checked quantitatively.

None of the lattice investigations of hadronic masses
at non-vanishing temperatures has attempted to carry out
the continuum limit.
Most of the mentioned analyses are based on the staggered
discretization. 
A straightforward
computation in the Wilson formulation of lattice QCD
and a comparison of the results with the staggered ones
would thus help to gain an idea about
the discretization effects.
This is the main goal of this paper.

Since we are extracting screening masses 
and Lorentz invariance is lost
at finite $T$ due to the heat bath, 
we have also computed spatial correlation functions
projected onto some non-vanishing momenta and on
the lowest non-vanishing bosonic Matsubara frequency.
The purpose of this attempt is to test for a sizeable
difference between spatial and temporal momentum
contributions to the vacuum polarization tensor.

As a by-product of the attempt to construct meson
operators with good overlap to the groundstate we also
were able to obtain information about the Bethe-Salpeter
amplitudes of the investigated mesons, the pion and the
$\rho$.

The paper is organized as follows: in the next section we present
some details of the simulation. This is followed by the
presentation and discussion of the obtained screening
masses in the pion and the $\rho$ channel, both below and above
the deconfinement transition.
In section \ref{sec:disp} we test the dispersion relation 
at non-vanishing temperature.
Section \ref{sec:wavefunc} contains our estimates of the wave functions.

\section{The simulation}
\label{sec:simu}

The results to be presented here are based on gauge field configurations
which have been generated with the standard Wilson gluon action.
We used a pseudo-heatbath algorithm \cite{CM}
with FHKP updating \cite{FHKP} in the $SU(2)$ subgroups. Each
heatbath iteration is supplemented by 4 overrelaxation
steps \cite{Adler}.
We have simulated at three values of the bare coupling,
$\beta = 6/g^2 = 6.0, 6.2$ and $6.4$.
At these $\beta$ values the lattice spacing has been determined
from quite a variety of observables.
Depending on the quantity, the results spread 
over a range of about 10 \% of the central value,
however, within the error bars, agreement is observed.
In order to obtain the physical temperature of the lattices
in units of the critical one, for definiteness we have 
consistently chosen to 
set the scale by the string tension \cite{Klassen},
$T/T_c = (T/\sqrt{\sigma}) \cdot (\sqrt{\sigma}/T_c)$,
taking $T_c/\sqrt{\sigma}$ from \cite{Beinlich}.
The temporal extent of the finite temperature lattices of
$N_\tau = 8$ then corresponds to physical temperatures
of $T = 0.93(1) T_c, 1.23(1) T_c$ and $1.63(2) T_c$ 
at $\beta = 6.0, 6.2 $ and $6.4$
respectively. The spatial volume of the lattices was chosen as 
$24^3$ and $32^3$ at $\beta = 6.0$,
$24^3$ at $\beta = 6.2$ and 
$24^2 \times 64$ at $\beta = 6.4$. 
In addition to the finite temperature simulations we have carried out
runs at zero temperature on lattices of size
$16^3 \times 32$ at $\beta = 6.0$ and $24^3 \times 48$ at
$6.2$ mainly in order to supplement the available literature
data on meson masses by results at quark masses in the range
between the strange and the charm quark mass.
We have generated between 50,000 and 80,000 gauge field
configurations and have analyzed configurations
separated by 500 or 2000 sweeps (see Table~\ref{runpara}).
Autocorrelations have been checked to be negligible.

For the fermion part of the action we used the ${\cal O}(a)$
Symanzik-improved Sheikho\-les\-lami-Wohlert action \cite{sheik} 
with a tree level clover coefficient of $c_{sw} = 1.0$.
The inversion of the Dirac matrix was carried out by means
of an overrelaxed MR algorithm \cite{Paige} at moderate to 
large quark mass values
and by the BiCGStabI \cite{Frommer} for light quarks.
In both cases an even-odd partitioning was employed.
We encountered only very few moderately exceptional configurations.
Including or omitting those 
did not modify the expectation values.
With the same combination of actions as in the present study,
the (latest) values of the critical hopping parameter at zero
temperature defined by the vanishing of the pion mass
have been determined as 
$\kappa_c = 0.14556(6)$ \cite {edin526} and
$\kappa_c = 0.14549(2)$ \cite {rom1157} at $\beta = 6.0$,
$\kappa_c = 0.14315(2)$ \cite {edin524} and
$\kappa_c = 0.14315(1)$ \cite {rom1157} at $\beta = 6.2$ as well as
$\kappa_c = 0.14143(3)$ \cite {rom1157} at $\beta = 6.4$.
These numbers may be used to convert the various values of the 
hopping parameter into an estimate of the corresponding
bare quark masses by means of
\beq
m_q a = \ln \left[ 1 + \frac{1}{2}\left(\frac{1}{\kappa}
-\frac{1}{\kappa_c}\right)\right]
\label{quarkmass}
\eeq
This has been used also in Table~\ref{runpara} where our
run parameters are summarized.

\begin{table}[!h]
  \begin{center}         
    \begin{tabular}[t]{|cc|ccc|cr|}
    \hline
    \multicolumn{7}{|c|}{ Simulation Parameters} \\
    \hline
$\beta$ & $a$ [fm] &
    \rule[-2mm]{0mm}{6.5mm} Lattice & $T/T_c$ & 
                       statistics & $\kappa$ & $m_q$ [MeV] \\
\hline
6.0 & 0.105 & $16^3 \times 32$        & 0    & ${\cal O}(100)$ & 0.141   &  200 \\
    &      &                         &      &                 & 0.130   &  750 \\
    &      &                         &      &                 & 0.128   &  800 \\ 
    &      & $24^3 \times 8 $        & 0.93  & ${\cal O}(100)$ & 0.141   &  200 \\
    &      &                         &      &                 & 0.130   &  750 \\
    &      &                         &      &                 & 0.128   &  800 \\
    &      & $32^3 \times 8 $        & 0.93  & ${\cal O}(40)$  & 0.145   &   30 \\
    &      &                         &      &                 & 0.1445  &   50 \\
    &      &                         &      &                 & 0.144   &   80 \\
    &      &                         &      &                 & 0.141   &  200 \\
\hline                                                                  
6.2 & 0.077 & $24^3 \times 48$        & 0    & ${\cal O}(100)$ & 0.141   &  150 \\
    &      &                         &      &                 & 0.130   &  800 \\
    &      &                         &      &                 & 0.128   &  950 \\
    &      & $24^3 \times 8 $        & 1.23  & ${\cal O}(30)$  & 0.1428  &   20 \\
    &      &                         &      &                 & 0.14232 &   60 \\
    &      &                         &      &                 & 0.14151 &  100 \\
    &      &                         &      &                 & 0.136   &  450 \\
    &      &                         &      &                 & 0.128   &  950 \\
\hline
6.4 & 0.058 & $24^2 \times 64 \times 8$ & 1.63 & ${\cal O}(100)$ & 0.1409  &   50 \\
    &      &                         &      &                 & 0.1406  &   80 \\
    &      &                         &      &                 & 0.1403  &  110 \\
    &      &                         &      &                 & 0.14    &  140 \\
    &      &                         &      &                 & 0.13    & 1100 \\
\hline
    \end{tabular}
  \end{center}
\caption{
\label{runpara}
         A summary of our run parameters. The physical value of the
         quark masses has been estimated by using eq.~(\ref{quarkmass})
         together with the physical value of the lattice spacing
         as obtained from string tension determinations. 
         For definiteness, $\sqrt{\sigma}$ has been taken
         as 420 MeV.
         }
\end{table}

From the computed quark propagators we constructed correlation
functions of operators with the quantum numbers of the
pseudoscalar and the vector meson.
To improve the projection onto the lowest
energy state, a gauge invariant extended operator was used
on the sink site \cite{Lacock}:
\beq
{\cal M}_R(x) = \sum_{\pm \vec e}{\overline \Psi}_\alpha^i (x)
\, {\cal U}^{i,j}(x \rightarrow x+R \vec e \,) \, \Gamma_{\alpha,\beta}
\, \Psi_\beta^j(x+R \vec e \,)
\label{operator}
\eeq
Here, $\Psi$ and ${\overline \Psi}$ are a quark and an antiquark field
separated by a distance $R$,
$i,j$ denote color indices and $\alpha, \beta$ are spinor ones.
Both indices are to be summed over. 
The quantum numbers are selected by choosing
$\Gamma = \gamma_5$ for the pseudoscalar and
$\Gamma = \gamma_\mu$ for the vector channel respectively.
In the vector channel we have averaged over the polarizations
$\mu$ perpendicular to the correlation direction.
The explicit sum in eq.~(\ref{operator}) is over 
all unit vectors $\vec e$ perpendicular
to the correlation direction.
This extended operator is made gauge invariant by introducing
the color parallel transporter ${\cal U}$ from $x$ to
$x + R \vec e$.
In order to further improve the projection and in an attempt
to resemble the gluon cloud \cite{Lacock} the parallel transporter
is built from smeared link fields.
At the source site a strictly local operator ($R = 0$) was
put on the lattice.

The most general correlation function $C_R(\vec p,t)$ is thus obtained as
\beqa
C_R(\vec p,t) & \; = \; & \sum_{\vec x} e^{- i \vec p \, \vec x}  
         \langle P_R(\vec x,t) \, P^\dagger (\vec 0,0) \rangle \nonumber \\
& \; = \; &  < 0 | P_R | P(\vec p) > < 0 | P | P(\vec p) >^* 
 \left\{ e^{ - E_P(\vec p) \, t} + 
         e^{ - E_P(\vec p) \, (N_\tau - t)}
    \right\} + \cdots
\label{corrfct}
\eeqa
where in this particular example the pseudoscalar correlation in the
temporal direction was chosen. 
In eq.~(\ref{corrfct}), the exponential fall-off is given by the
energy $E_P(\vec p)$ of the state $| P(\vec p) >$ at momentum
$\vec p$. The dots indicate contributions from excited states
with the right quantum numbers.
At non-vanishing temperature, because
of the limited extent of the lattices in the temporal direction,
$t \leq 1/T$, we computed spatial correlation functions in the
$z$ direction,
$C_R({\vec {\tilde p}},z)$, where ${\vec {\tilde p}}$ denote
the momentum components perpendicular to the $z$ direction,
${\vec {\tilde p}}=(p_x,p_y,p_t)$.

\begin{figure}[t]
  \begin{center}
    \epsfig{file=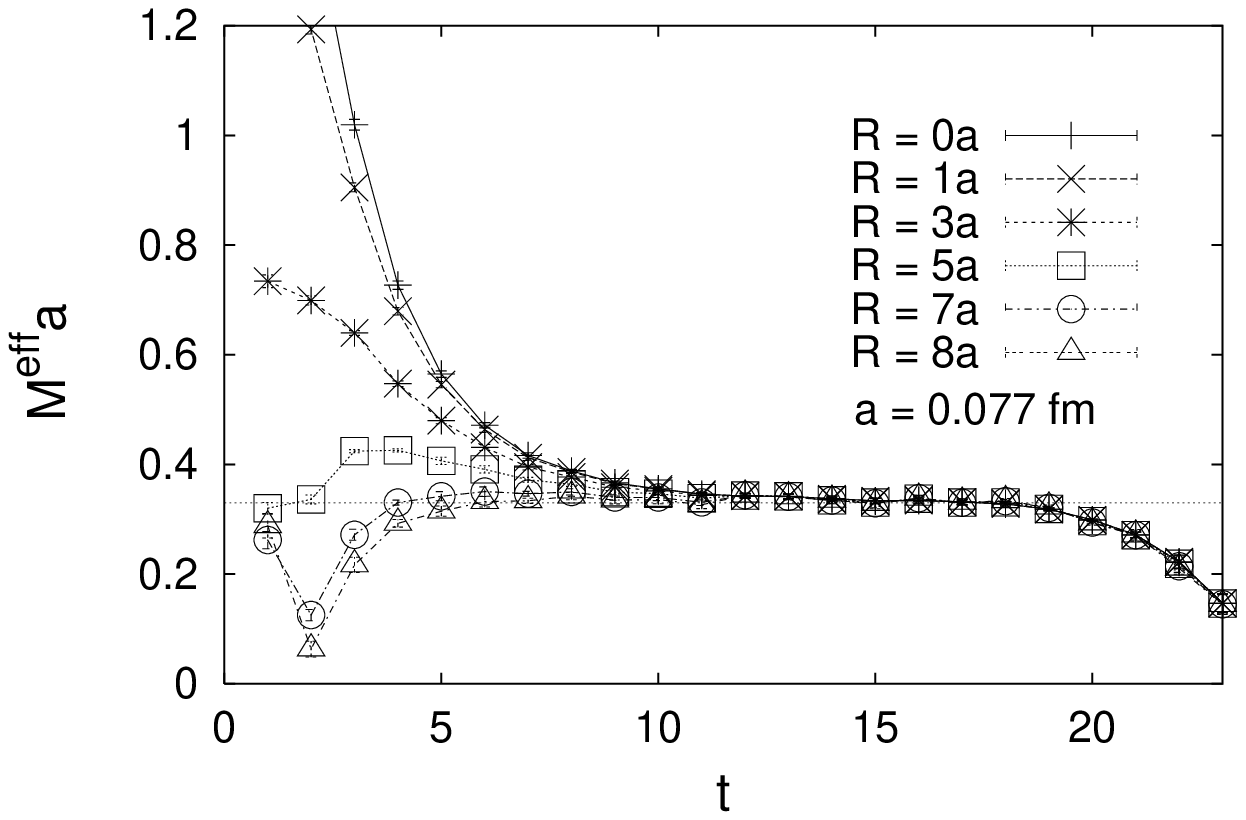,width=114mm}
    \epsfig{file=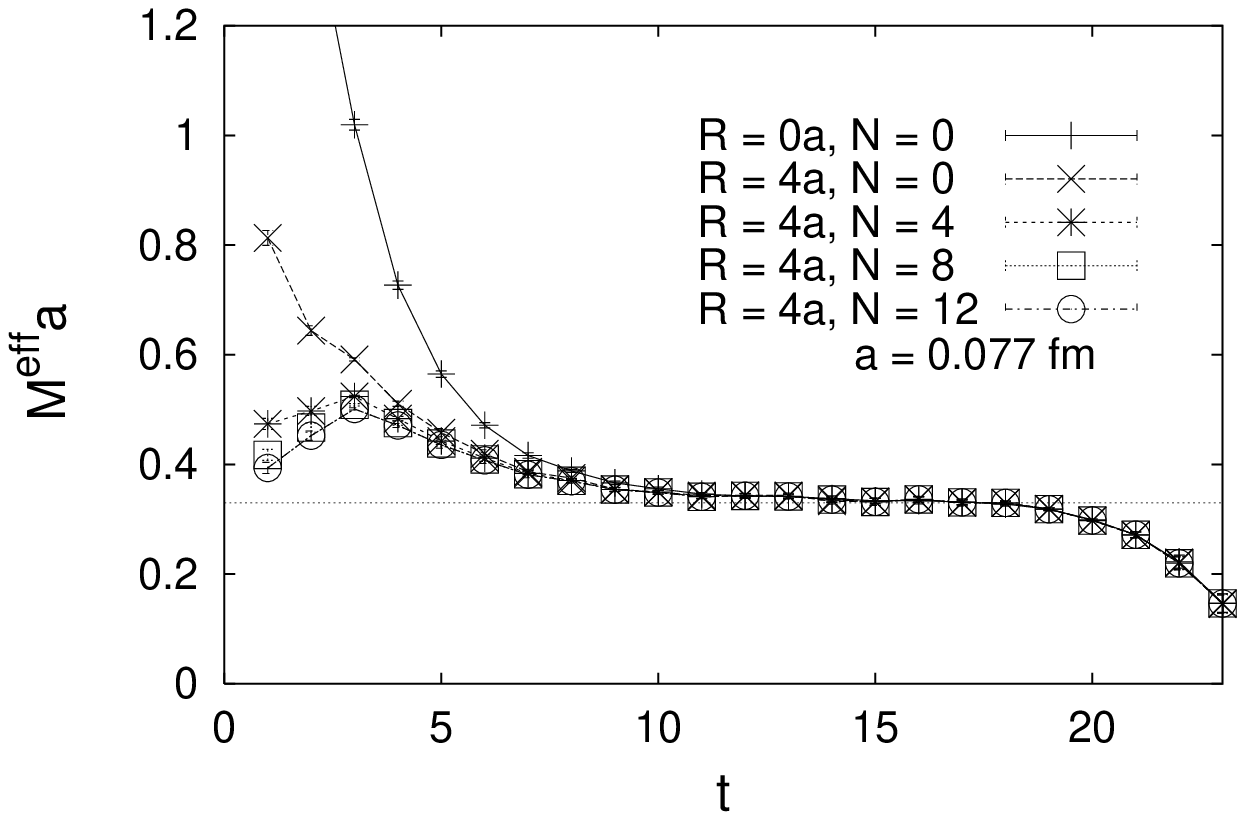,width=114mm}
    \caption{
\label{projection}
             The effective mass $M^{\rm eff}(t)$ 
             as a function of $t$. 
             The example shows the pseudoscalar channel at zero
             temperature at $\beta=6.2$ and $\kappa=0.141$.
             The upper figure demonstrates the effect of varying
             the distance $R$ between quark and antiquark while
             the number of smearing iterations was fixed to 4 in
             this example.
             In the lower plot, $R$ is fixed at 4 and the number of 
             smearing iterations is varied between 0 and 12.
             }
  \end{center}
\end{figure}

The improvement procedure described above leaves quite some freedom
in choosing optimal parameters. 
As for the distance $R$ between
quark and antiquark, on each configuration we have computed
the correlation functions for a variety of different $R$ values. 
This allowed us to find the optimal separation for each lattice 
spacing and temperature individually. 
An example of how the separation $R$ changes the projection to
the lowest mass state is shown in Figure~\ref{projection}.
Here we plot the effective mass, defined as
$M^{\rm eff}(t) = \ln\{C_R(t)/C_R(t+1)\}$ for $\vec p = \vec 0$,
of the pseudoscalar 
as a function of $t$ for various $R$ values. The data has been obtained
on the zero temperature lattice at $\beta = 6.2$ at $\kappa = 0.141$.
The plot illustrates that the contribution of excited states becomes
considerably smaller when $R$ is raised from 1 to 5 lattice spacings
in this example.
The data flattens off and reaches a plateau at smaller time 
separations between source and sink. This allows to extract the 
lowest mass much more reliably.
At $R \geq 7$ the large $t$ limit is approached from below as
a single term contributing to eq.~(\ref{corrfct}) is not
positive-definite. At the rightmost data point in the figure
the effective mass drops slightly because the periodicity of
the lattice is being felt by the correlation function.
When $\beta$ is varied we observe that the optimal distance
in lattice units approximately
scales with the lattice spacing i.e. stays constant
in physical units.

Contrary to the strong $R$ dependence of the effective masses
which appears to be physical, details of the gauge field
smearing procedure do not seem to matter so much.
We have adopted the APE prescription \cite{APEsmear}
with a weight of 2 for the link term and 1 for the contribution
of the staples. As noted also in \cite{Lacock},
the precise value of the ratio is not too important,
contrary to the case of Wilson loops.
The smeared links were projected back to
$SU(3)$ elements. In test runs it turned out that varying the 
number of smearing iterations between 4 and 12 does not have 
a big impact
on the length of the plateau in effective mass plots,
see Figure~\ref{projection}.
Moreover, between the $\beta$ values analyzed we did not
observe significant differences.

\section{Masses}
\label{sec:masses}

In this section we present our results on the masses. These
were obtained from correlation functions 
$C_R(\vec p = 0,t)$, eq.~(\ref{corrfct}), at zero momentum
in which case the exponential fall-off is given by the mass,
$E_H(\vec p = 0) = M_H$.
For spatial correlation functions the exponential fall-off
defines the screening mass.

In order to obtain an estimate of the mass of the
lowest state contributing to a given correlation function
we first compared the effective mass plots, 
e.g. Figure~\ref{projection},
for various quark pair distances $R$ in search of the
optimal $R$ value with regard to the 
onset and stability of the plateau. Subsequently, 
at the chosen $R$ value 
fits over intervals $[t_{\rm min}, N_\tau-t_{\rm min}]$
with varying $t_{\rm min}$ 
(similarly for the spatial correlations at $T \neq 0$)
were performed, again checking for
stability of the mass value.
Likewise, we symmetrized the correlation functions around
midpoint, $N_\tau/2$ and $N_\sigma/2$ respectively, and
carried out fits including the full covariance matrix.
Again, the minimum separation from the source was varied.
The results quoted, mass values as well as errors, 
are obtained from the latter fits, selecting
the fit interval by the best $\chi^2$ value. The errors given
include an estimate of the systematic error as suggested by
a remaining dependence of the mass on $R$ and $t_{\rm min}$.
Finally, also
two-state fits were applied to correlation functions
including data at next-to-optimal $R$ values and smaller
separations from the source in order to further check for
consistency.

\begin{figure}[h!]
  \begin{center}
    \epsfig{file=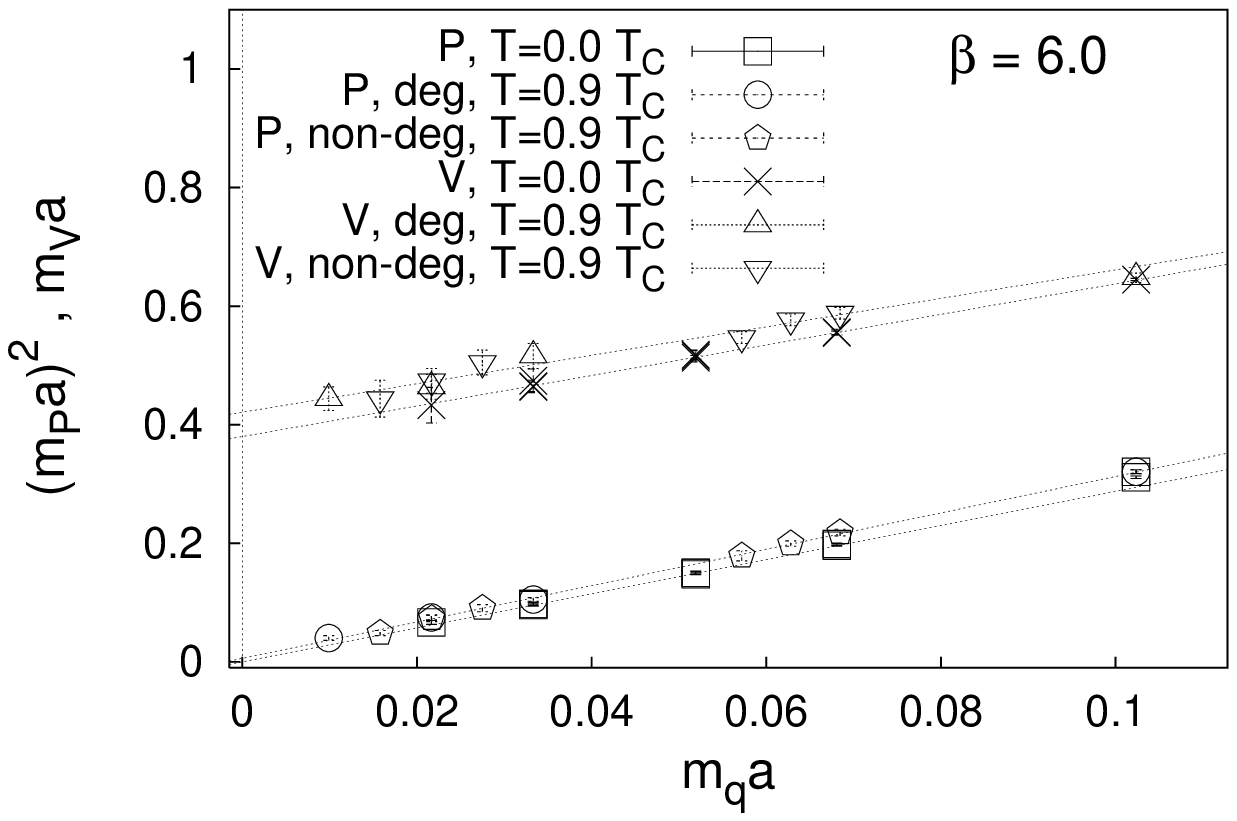,width=120mm}
    \epsfig{file=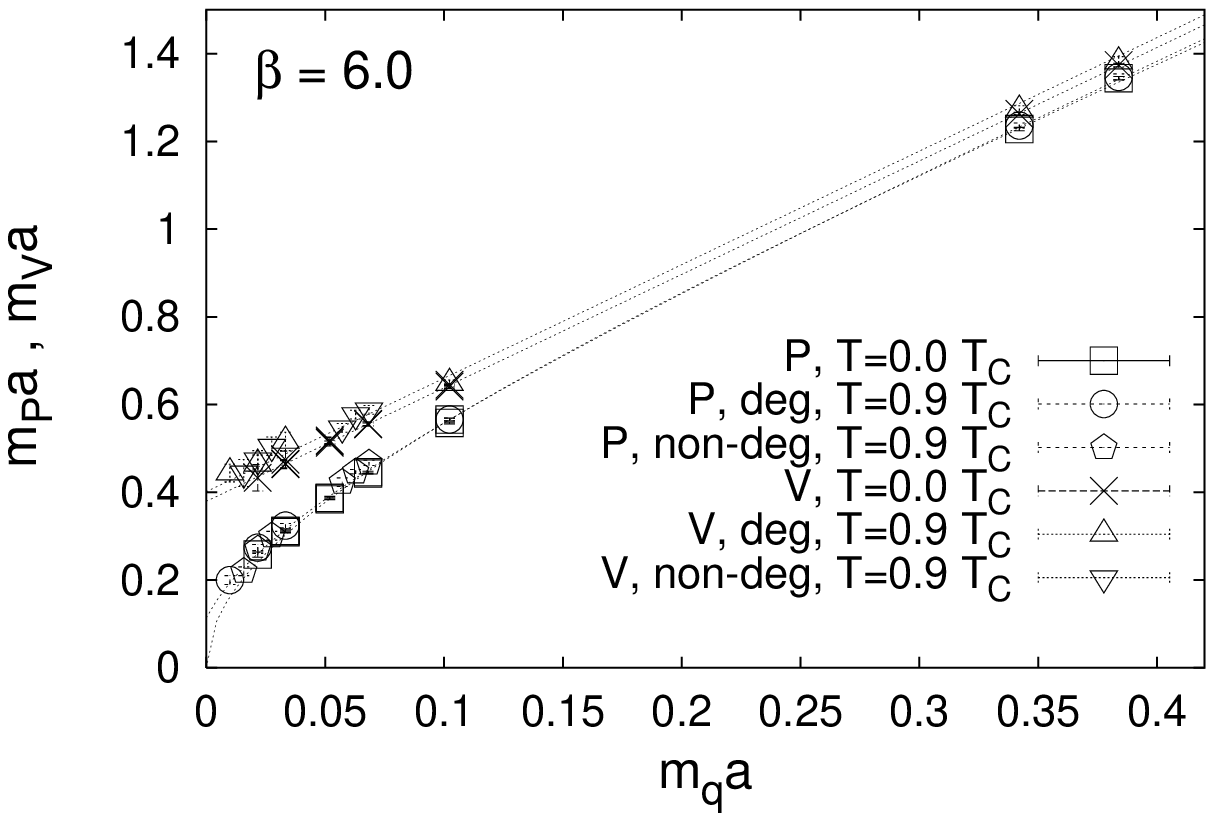,width=120mm}
    \caption{Meson masses at $T \simeq 0.93 \, T_c$ compared
             with the zero temperature results
             as a function of the quark mass.
             The zero-temperature data is partially
             taken from the literature 
             \protect\cite{edin526,rom1157,APEsmall}.
             The upper plot features the vectormeson mass
             and the pseudoscalar mass squared
             in the vicinity of the chiral limit
             while in the lower part the data for the
             entire quark mass range are shown. Here both 
             masses are plotted linearly. The lines are fit results
             explained in the text.
    \label{mass09}
    }
  \end{center}
\end{figure}

\begin{figure}[h]
  \begin{center}
    \epsfig{file=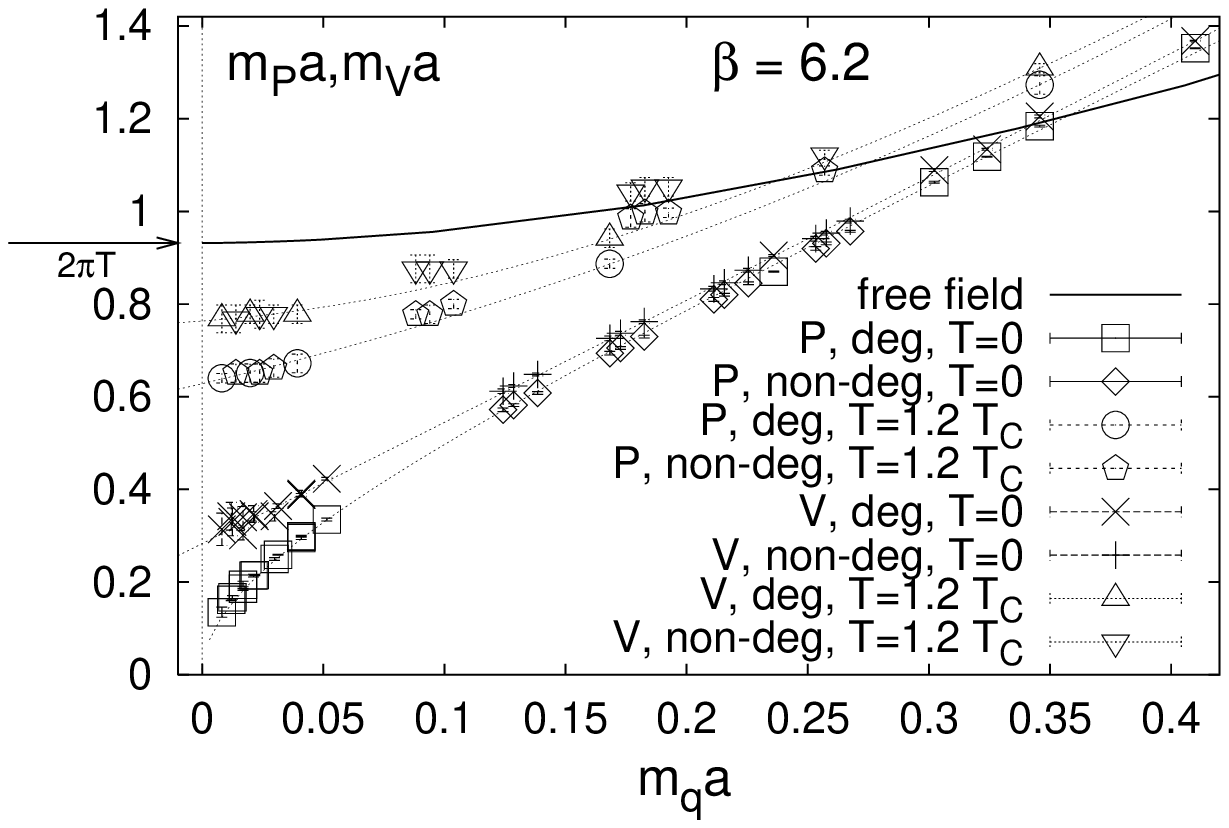,width=114mm}\\
    \epsfig{file=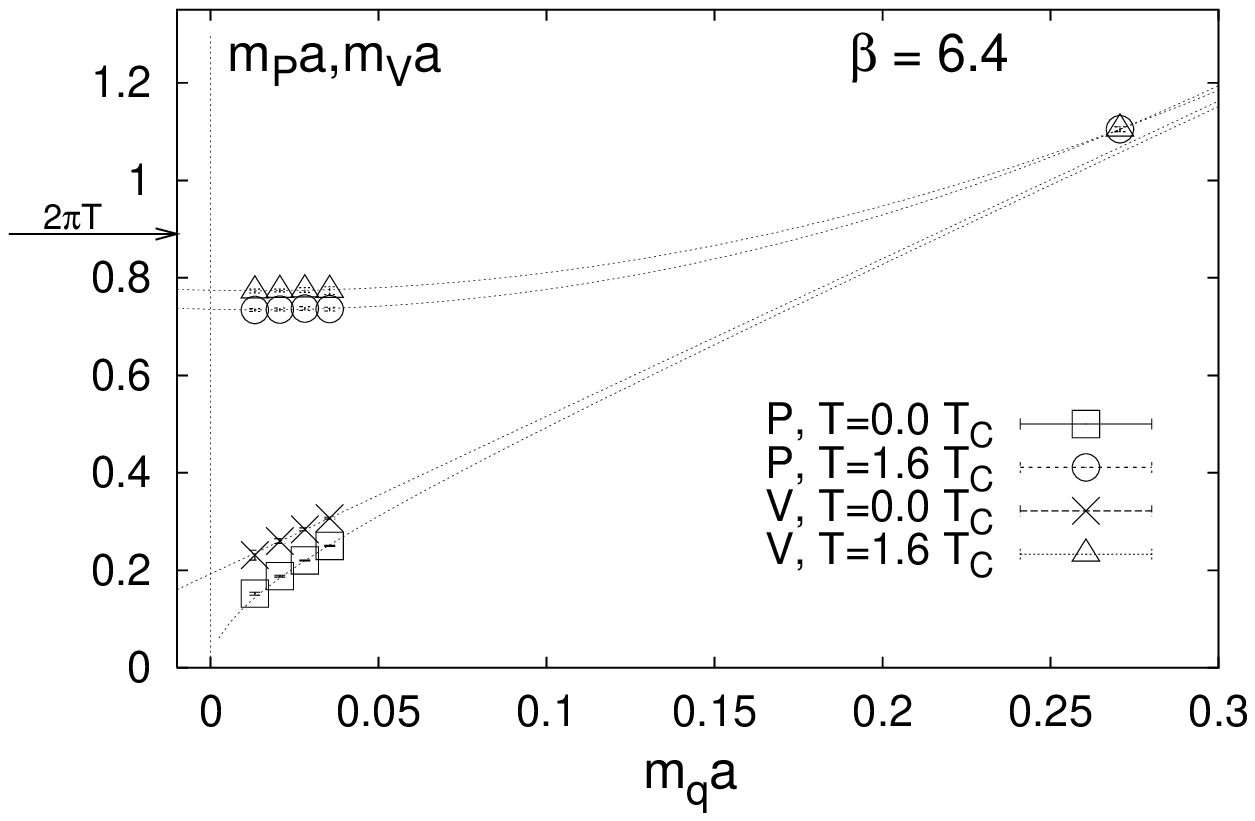,width=114mm}\\
    \caption{Meson masses at $T \simeq 1.23 T_c$ and
             $1.63 T_c$ compared with zero temperature data
             partially taken from the literature 
             \protect\cite{edin526,rom1157,APEsmall}.
    \label{mass12}
    }
  \end{center}
\end{figure}

The results for the ground state masses in the
pseudoscalar and the vector channel at the temperatures
investigated are summarized in Tables \ref{mass60t} 
to \ref{mass64t} and 
are compared with the available zero temperature data
in Figures \ref{mass09} and \ref{mass12}. 
In the figures we have plotted the so-called pole mass
\beq
m_H = 2 \; {\rm sinh}\left(\frac{M_H}{2}\right)~.
\label{polemass}
\eeq
Eq.~(\ref{polemass}) arises from a lattice meson action with
a nearest neighbor symmetric difference discretization which
appears to be favored by studies of lattice dispersion
relations \cite{Rajan,rom1157}.
Using $m_H$ instead of the coefficient $M_H$ of the exponential
fall-off partly corrects for ${\cal O}(a m)$ lattice artefacts.
We observe that $m_V$ for the vectormeson at zero
temperature, unlike $M_V$, is strikingly linear in the
quark mass up to values in the vicinity of the charm quark mass.
For small meson masses the difference between $m_H$ and $M_H$
is of course negligible.

\begin{table}
\begin{minipage}[t]{72mm}



\begin{tabular}{|c|c|c|c|}\hline
\multicolumn{4}{|c|}{\rule[-3mm]{0mm}{7mm} \boldmath $16^{3} \times 32 , \;
  \beta  \! = \!  6.0, \; T \! = \! 0$ \unboldmath} \\
\hline
\rule[-3mm]{0mm}{7mm} $\kappa_{1}$ & $\kappa_{2}$ & $M_{P}$ & $M_{V}$ \\ \hline
0.1280      & 0.1280      & 1.259 (3) & 1.284 (4)     \\
0.1300      & 0.1300      & 1.162 (3) & 1.191 (3)     \\
0.1410      & 0.1410      & 0.559 (3) & 0.630 (4)     \\
\hline 
\end{tabular}\\

\vspace*{1mm}
\begin{tabular}{|c|c|c|c|}\hline
\multicolumn{4}{|c|}{\rule[-3mm]{0mm}{7mm} \boldmath $24^{3} \times 8 , \;
  \beta  \! = \!  6.0, \; T \! = \! 0.93 \, T_{c}$ \unboldmath} \\
\hline
\rule[-3mm]{0mm}{7mm} $\kappa_{1}$ & $\kappa_{2}$ & $M_{P}$ & $M_{V}$ \\ \hline
0.1280      & 0.1280      & 1.262 (7) & 1.29  (1)     \\
0.1300      & 0.1300      & 1.168 (6) & 1.197 (8)     \\
0.1410      & 0.1410      & 0.557 (8) & 0.64  (1)     \\
\hline 
\end{tabular}\\

\vspace*{1mm}
\begin{tabular}{|c|c|l|l|}\hline
\multicolumn{4}{|c|}{\rule[-3mm]{0mm}{7mm} \boldmath $32^{3} \times 8 , \; \beta
   \! = \!  6.0,  \; T \! = \! 0.93\, T_{c}$ \unboldmath} \\
\hline
\rule[-3mm]{0mm}{7mm} $\kappa_{1}$ & $\kappa_{2}$ & $M_{P}$ &$M_{V}$          \\ \hline
                                   & 0.14100      & 0.559 (5) & 0.64  (1)     \\ 
                                   & 0.14400      & 0.463 (6) & 0.59  (1)     \\ 
\raisebox{1.5ex}[-1.5ex]{0.1410}  & 0.14450      & 0.443 (5) & 0.57  (1)     \\ 
                                   & 0.14500      & 0.42  (1) & 0.54  (1)     \\ 
\hline
\hline
                                   & 0.14400      & 0.323 (5) & 0.51  (2)     \\ 
0.1440                            & 0.14450      & 0.300 (1) & 0.50  (2)     \\ 
                                   & 0.14500      & 0.27  (1) & 0.47  (2)     \\ 
\hline
\hline
                                   & 0.14450      & 0.273 (7) & 0.46  (2)     \\
\raisebox{1.5ex}[-1.5ex]{0.1445}  & 0.14500      & 0.22  (1) & 0.44  (3)     \\
\hline
\hline
0.1450                            & 0.14500      & 0.20  (1) & 0.44  (2)     \\
\hline 
\end{tabular}
\vspace*{-1.2mm}
%
%
\caption[$M_{H}$ at $\beta  \! = \!  6.0$ for  $T \! = \! 0$ and $T \! = \! 0.93 \; T_{c}$]{\label{mass60t}
  {\sl
    Meson masses $M_{H}$ of the pseudoscalar and vector meson 
    at $\beta  \! = \!  6.0$, $T  \! = \!  0$ on a $16^{3} \! \times \! 32$ lattice
    and $T  \! = \!  0.93\; T_{c}$ on $24^{3} \! \times \! 8$ and
    $32^{3} \! \times \! 8$ lattices.}
  }
%
\end{minipage} \hfill
\begin{minipage}[t]{72mm}


\begin{tabular}{|c|c|c|c|}\hline
\multicolumn{4}{|c|}{\rule[-3mm]{0mm}{7mm} \boldmath $24^{3} \times 48 , \;
  \beta  \! = \!  6.2, \; T \! = \! 0$ \unboldmath} \\
\hline
\rule[-3mm]{0mm}{7mm} $\kappa_{1}$ & $\kappa_{2}$ & $M_{P}$ & $M_{V}$ \\ \hline
\rule[0mm]{0mm}{3.5mm}0.1280      & 0.1280      & 1.124 (2) & 1.141 (2)     \\
0.1300      & 0.1300      & 1.018 (2) & 1.040 (2)     \\
0.1410      & 0.1410      & 0.333 (2) & 0.420 (4)     \\
\hline 
\end{tabular}\\

\vspace*{2mm}
\begin{tabular}{|c|c|c|c|}\hline
\multicolumn{4}{|c|}{\rule[-3mm]{0mm}{7mm} \boldmath $24^{3} \times 8 , \;
  \beta  \! = \!  6.2, \; T \! = \! 1.23 \, T_{c}$ \unboldmath} \\
\hline
\rule[-3mm]{0mm}{7mm} $\kappa_{1}$ & $\kappa_{2}$ & $M_{P}$ & $M_{V}$ \\ \hline
\rule[0mm]{0mm}{3.5mm}0.1280      & 0.1280      & 1.110 (4) & 1.110 (5)     \\
0.1300      & 0.1300      & 0.995 (5) & 1.011 (4)     \\
0.1410      & 0.1410      & 0.350 (4) & 0.410 (8)     \\
\hline 
\end{tabular}\\

\vspace*{2mm}
\begin{tabular}{|c|c|c|c|}\hline
\multicolumn{4}{|c|}{\rule[-3mm]{0mm}{7mm} \boldmath $24^{3} \times 8 , \; \beta
   \! = \!  6.2,  \; T \! = \! 1.23\, T_{c}$ \unboldmath} \\
\hline
\rule[-3mm]{0mm}{7mm} $\kappa_{1}$ & $\kappa_{2}$ & $M_{P}$ &$M_{V}$\\ \hline
\rule[0mm]{0mm}{3.5mm}            & 0.12800 & 1.20 (2) & 1.23  (2)  \\
                                  & 0.13600 & 1.04 (1) & 1.07  (1)  \\
0.12800                           & 0.14151 & 0.96 (1) & 1.01  (2)  \\
                                  & 0.14232 & 0.96 (2) & 1.01  (2)  \\
                                  & 0.14280 & 0.95 (2) & 1.00  (2)  \\ 
\hline                                                             
\hline                                                             
\rule[0mm]{0mm}{3.5mm}            & 0.13600 & 0.86 (1) & 0.91  (2)  \\
                                  & 0.14151 & 0.78 (1) & 0.85  (2)  \\
\raisebox{1.5ex}[-1.5ex]{0.13600} & 0.14232 & 0.76 (2) & 0.85  (3)  \\
                                  & 0.14280 & 0.76 (1) & 0.85  (3)  \\
\hline                                                             
\hline                                                             
\rule[0mm]{0mm}{3.5mm}            & 0.14151 & 0.66 (2) & 0.76  (2)  \\
0.14151                           & 0.14232 & 0.65 (2) & 0.76  (2)  \\
                                  & 0.14280 & 0.64 (2) & 0.76  (3)  \\
\hline                                                             
\hline                                                             
\rule[0mm]{0mm}{3.5mm}            & 0.14232 & 0.64 (2) & 0.76  (2)  \\
\raisebox{1.5ex}[-1.5ex]{0.14232} & 0.14280 & 0.64 (2) & 0.75  (3)  \\
\hline                                                             
\hline                                                             
\rule[0mm]{0mm}{3.5mm}0.14280     & 0.14280 & 0.63 (1) & 0.75  (3)  \\
\hline 
\end{tabular}\\

\vspace*{-1.2mm}
%
%
\caption[$M_{H}$ at $\beta  \! = \!  6.2$ for  $T \! = \! 0$, $T \! = \! 1.23 \; T_{c}$ and $T \! = \! 1.63 \; T_{c}$]{\label{mass62t}
  {\sl
    Meson masses $M_{H}$ of the pseudoscalar and vector meson 
    at $\beta  \! = \!  6.2$, $T  \! = \!  0$ on a $24^{3} \! \times \! 48$ lattice
    and $T  \! = \!  1.23\; T_{c}$ on a $24^{3} \! \times \! 8$ lattice.}
  }
\end{minipage}
\end{table}

\begin{table}
\begin{center}
\begin{minipage}[t]{80mm}
\begin{tabular}{|c|c|c|c|}\hline
\multicolumn{4}{|c|}{\rule[-3mm]{0mm}{7mm} \boldmath $24^{2} \times 64 \times 8 , \;
  \beta  \! = \!  6.4, \; T \! = \! 1.63 \, T_{c}$ \unboldmath} \\
\hline
\rule[-3mm]{0mm}{7mm} $\kappa_{1}$ & $\kappa_{2}$ & $M_{P}$ & $M_{V}$ \\ \hline
\rule[0mm]{0mm}{3.5mm}0.13000      & 0.13000      & 1.055 (5) & 1.085 (2)     \\
0.14000      & 0.14000      & 0.720 (4) & 0.756 (7)     \\
0.14030      & 0.14030      & 0.721 (4) & 0.757 (4)     \\
0.14060      & 0.14060      & 0.719 (3) & 0.756 (3)     \\
0.14090      & 0.14090      & 0.718 (3) & 0.755 (4)     \\
\hline 
\end{tabular}
%
%
\caption[$M_{H}$ at $\beta  \! = \!  6.2$ for  $T \! = \! 0$, $T \! = \! 1.23 \; T_{c}$ and $T \! = \! 1.63 \; T_{c}$]{\label{mass64t}
  {\sl
    Meson masses $M_{H}$ of the pseudoscalar and vector meson 
    at $\beta  \! = \!  6.4$, $T  \! = \!  1.63\; T_{c}$ on a $24^{2} \! \times \! 64 \! \times \! 8$ lattice.}
  }
\end{minipage}
\end{center}
\end{table}

Figure~\ref{mass09} summarizes the results 
at $\beta = 6.0$
which on the high temperature lattice corresponds to a
temperature of $0.93 \, T_c$.
The upper part of the figure shows $m_V$ and $m_P^2$
at small quark mass values while the lower part covers the entire
quark mass range explored.
The meson masses are plotted as function of $m_q$ where
the quark mass is obtained from eq.~(\ref{quarkmass})
with $\kappa_c\,(T\!=\!0)$ determined from a linear fit in $m_q$
to the combined zero temperature literature data 
on $m_P^2$ \cite{edin526,rom1157}
up to quark mass values $m_q a < 0.08$. 
As can be seen from the lowest lying line drawn in the upper part 
of Figure~\ref{mass09},
the linear fit works well and deviations begin
to emerge at $m_q a \simeq 0.1$.
Our result for the critical kappa value is
$\kappa_c\,(T\!=\!0) \!=\! 0.14542(2)$ which
deviates marginally from the values quoted in
\cite{edin526,rom1157,APEsmall}
because we have chosen a fit procedure slightly different from the
one adopted there. 
Note however, that our value is well within the spread of
the quoted $\kappa_c$ values.
Moreover, the upper part of Figure~\ref{mass09} shows that
the finite temperature pion (screening) mass 
retains a small but non-vanishing value at $\kappa_c\,(T=0)$.
A fit to the $m_P^2$ data at quark mass values up to 0.08
of the form $c + s \, m_q$,
shown as the second lowest dotted line,
leads to an intercept
$c = 0.006(3)$ at $m_q= 0$. The slope $s=3.07(8)$ is 
a little larger than the value of $s=2.87(3)$ obtained
from the equivalent fit to the $T=0$ data.
Alternatively, one can perform a fit with a 
temperature dependent $\kappa_c$ leading to a slightly
different value, $\kappa_c\,(T) = 0.14550(4)$.

For the vectormeson masses,
when plotted as a function of the zero temperature quark mass,
we find that, as a whole, they apparently tend to be
somewhat larger than the zero temperature values. In both cases,
fits linear in the quark mass work well and
we obtain $m_V = 0.42(2)$ at $T \simeq 0.93 \, T_c$
as opposed to 0.38(1) at $T = 0$ 
in the chiral limit.
This is shown as the appropriate lines in the upper part
of Figure~\ref{mass09}.
At least part of this difference could, however, be absorbed into the
shift in $\kappa_c$ mentioned above. 

In the lower part of Figure~\ref{mass09}
$m_V$ and $m_P$ are shown linearly
over the entire quark mass range explored.
At zero temperature, the vectormeson mass is linear in
$m_q$ over the entire range. 
A fit ansatz including a term quadratic in the quark mass 
returns a value of $0.05(19)$ for its coefficient
which is compatible with 0.
The intercept is obtained as $0.381(6)$.
Within errors this intercept is in agreement with the
result of a linear fit,
$m_V = 0.379(2) + 2.588(5) \, m_q$ which is shown in the figure.
Note, that these numbers are in agreement with the results
for the intercept obtained from the fit to the small quark mass data only.
The mass of the pseudoscalar shows a behavior proportional to
$\sqrt{m_q}$ at small
quark masses, as expected from chiral perturbation theory,
which turns into a linear one at larger $m_q$ values.
Correspondingly, we chose a fit ansatz of the form
\beq
m_P = \sqrt{ \, b \, m_q + d \, m_q^2} 
\label{pionfit}
\eeq
which returns $b=2.57(1)$ and $d=5.39(6)$, shown as the
lowest line in the figure.

Regarding the results at non-vanishing temperature,
also at large quark masses we do not
observe significant differences to the zero temperature data. 
However, the meson mass values are again plotted at the
zero temperature quark masses,
i.e. as a function of 
$1/\kappa - 1/\kappa_c\,(T\!=\!0)$ with our value of 
$\kappa_c$ at $T\!=\!0$.
In a fit to the pseudoscalar mass data over the entire $m_q$ range,
we therefore allow for a non-vanishing chiral limit and replace $m_q$ by
$m_q \rightarrow m_q + c$ in the ansatz eq.~(\ref{pionfit}).
This fit leads to the second lowest dotted curve 
in the lower part of Figure~\ref{mass09}
with $c= 0.005(2)$, $b=2.3(1)$ and $d=5.9(2)$.
When fitting the vectormeson over all quark masses we observe
a slightly better performance in terms of $\chi^{2}$
of a fit including terms quadratic
in $m_q$,
$m_V = 0.413(6) + 2.45(5) \, m_q + 0.13(5) \, m_q^2$,
over a linear one,
$m_V = 0.401(5) + 2.59(1) \, m_q$.
The quadratic term is not large though and the intercepts have their
error bars touching each other. Note however that the intercept
of the quadratic fit is in slightly better accord with the
result of the linear fit to the small quark mass data.
The corresponding line in Figure~\ref{mass09} is the quadratic
function.

Summarizing the results at $T \simeq 0.93 \, T_c$, we note
that the pion (screening) mass exhibits only a very small value
in the zero temperature chiral limit which could be absorbed
by a slight shift in the critical kappa value.
This might also be a finite size effect as the finite temperature
lattice is somewhat bigger in spatial volume than the zero temperature
ones.
A non-vanishing temperature effect is perhaps seen in the
vectormeson data which shows an enhancement of 7 to 10 \% over
the zero temperature data. It remains to be seen, however, 
whether this finding persists in the continuum limit. On the other
hand, our results in the Wilson discretization are in good agreement
with observations made in earlier studies using 
staggered fermions \cite{Boyd1}.

The results for the meson masses at $\beta = 6.2$ and $6.4$
corresponding to the two temperatures above $T_c$ which are
investigated here, $T \simeq 1.23 \, T_c$ and $1.63 \, T_c$ 
respectively, are summarized in Figure~\ref{mass12}.
Again, we show $m_V$ and $m_P$ as a function of the quark mass
$m_q$ with $\kappa_c\,(T=0)$ determined as explained above 
from fits to the combined
zero temperature pion mass results \cite{rom1157,edin524}
at small quark mass.
The resulting critical $\kappa$ values are
$\kappa_c = 0.14313(2)$ and $0.14141(3)$ at 6.2 and 6.4, respectively.

The zero temperature data shows the same behavior as at
$\beta = 6.0$, including the linearity in $m_q$ of the vector
meson mass up to the largest quark masses explored.
At $\beta = 6.2$ we fitted $m_V$ at $T = 0$
over the entire quark mass range
both, quadratically in $m_q$, leading to
$m_V = 0.283(2) + 2.61(2) \, m_q + 0.10(4) \, m_q^2$ 
as well as linearly,
$m_V = 0.279(1) + 2.65(5) \, m_q$.
Again, the quadratic term was found to be small and the intercepts 
are stable. Using only data close to the chiral limit 
leads to an extrapolated value of
$m_\rho = 0.267(8)$, so that the deviation 
from the fits to the full data set is a little larger 
than at $\beta = 6.0$.
The dotted lines given in Figure~\ref{mass12} are the quadratic fits.
Note, that the data points labelled by
``{\tt non-deg}'' indicate meson masses obtained from propagators
with two quarks of non-degenerate masses. They are plotted as 
a function of an ''effective'' quark mass 
$\overline{m}_q = (m_{q,1} + m_{q,2})/2$ and fit very well
on the quadratic curves at $T=0$.
The pseudoscalar data was again fitted with the ansatz 
of eq.~(\ref{pionfit}) which gives
$b=1.846(8)$ and $d=6.22(3)$ in this case. This is shown again 
as the lowest line in the figure.

At $\beta = 6.4$ only data at small quark masses were available 
\cite{rom1157} and we have carried out a linear fit to the vector
meson, with the result
$m_V = 0.19(1) + 3.23(3) \, m_q $. 
Although $\kappa_c\,(T\!=\!0)$ was obtained from fits to $m_P^2$
in the vicinity of the chiral limit, in Figure~\ref{mass12} we
only show, as lines to guide the eye, the results of fits with 
eq.~(\ref{pionfit})
under the condition that $m_P \,\lsim\, m_V$ for large $m_q$. 

In contrast to the situation at $T < T_c$, at both temperatures
above the transition point the screening masses are
markedly different from the zero temperature results, 
especially at small quark mass values.
The curves interpolating the non-zero temperature data
are polynomials of third degree for the vectormeson at $T\!=\!1.23T_c$
and of second degree elsewhere. We have chosen this ansatz
because above $T_c$ no particular functional form is known so far. 
The curves are the results of fits to data for degenerate
quark masses and are mainly meant to guide the eye. 
The intercepts were obtained from these fits as
$m_\pi(1.23T_c)  = 0.63(2)$,
$m_\rho(1.23T_c) = 0.74(2)$ and
$m_\pi(1.63T_c)  = 0.735(5)$,
$m_\rho(1.63T_c) = 0.774(8)$ at the two temperatures
respectively. They are discussed in the following.

At high temperatures one expects that the plasma consists
of only weakly interacting quarks and gluons.
Correspondingly, the correlation functions with mesonic
quantum numbers should be described to first approximation
by the free propagation of a quark-antiquark pair.
In this case the exponential fall-off of the correlation functions
is not dominated at large distances by a single mass originating
from an isolated pole in the spectral function
but rather by all possible energies of quark and antiquark with
opposite ``momenta'' of equal size adding to zero total momentum
to which the correlation functions are projected on.
Correspondingly, effective masses taken from point to point
correlations ($R=0$ in eq.~(\ref{operator}))
exhibit a rather marked curvature
as function of the separation between the meson operators.
Tuning $R$ away from 0 leads to a suppression of higher
quark momenta in the momentum sums. This way, it is
possible to obtain plateau-like behavior in effective mass plots
also in the free quark case.
The lowest value of quark ``momentum'' in the temporal direction,
i.e. the smallest Matsubara frequency is $\pi T$ due to the
antiperiodic boundary condition for fermions in the $t$
direction. This leads to the lowest ``energy'' contributing
to spatial correlation functions of mesonic operators of
\beq
m_{H} = 2 \sqrt{ m_{q}^{2} + (\pi T)^{2}}
\label{freequarks}
\eeq
in the continuum. On a finite lattice, eq.~(\ref{freequarks})
is subject to finite volume and finite lattice spacing 
corrections, see e.g. \cite{MTc}.
In order to compare our simulation results at $T > T_c$ with
the free quark case, we have computed the free quark propagator
on a finite lattice analytically and performed the
sum over all lattice ``momenta'' $\vec{\tilde p}$ of the quarks
numerically. For definiteness, we then plot the effective
mass for point to point correlations 
at a separation $z = N_\sigma/4$. This results in the
solid curve shown in the upper part of Figure~\ref{mass12}.

As in the free case, in our results from the Monte Carlo simulations 
we are also able to 
identify plateaux in the effective mass plots
when the separation $R$, eq.~(\ref{operator}), is tuned. 
The vectormeson and, to a somewhat less degree,
the pseudoscalar
masses approach the free quark limit as is seen from
Figure~\ref{mass12}.
The remaining differences are becoming smaller with rising 
temperature.
Moreover, the fact that the masses for non-degenerate
quark combinations, plotted at 
$\overline{m}_q = (m_{q,1} + m_{q,2})/2$,
are above the ones for degenerate quarks would fit into this
picture as 
$2 \sqrt{\overline{m}_q^2 + (\pi T)^2} <
\sqrt{m_{q,1}^2 + (\pi T)^2} +
\sqrt{m_{q,2}^2 + (\pi T)^2}$.
At large quark masses the Monte Carlo data exceed the free quark curve.
This could be explained by noting that 
in the lattice version of eq.~(\ref{freequarks})
we have used the bare quark mass.
At the temperatures investigated one should presumably
compare with an effective quark mass which also accounts
for a thermal contribution \cite{Boyd3}.
In addition, it has been argued in the context of
dimensional reduction that the (confining) potential
of the reduced three-dimensional theory leads to
modifications of eq.~(\ref{freequarks}) for
screening masses \cite{Hansson}.
These modifications will be positive and 
of order $\sqrt{\sigma_{\rm spat}}$,
where $\sigma_{\rm spat}$ is the string tension of
spatial Wilson loops, and can be estimated to amount
to an ${\cal O}( 10\%)$ effect \cite{Martin}.

\begin{figure}[h]
  \begin{center}
    \epsfig{file=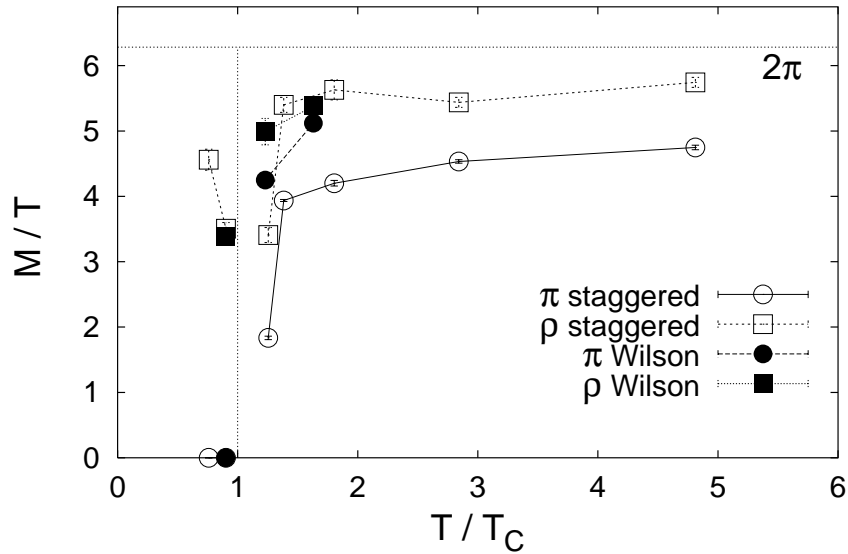,width=114mm}\\
    \caption{
             \label{stagg}
             Comparison with the results of quenched staggered
             simulations below \cite{Boyd1} and above $T_c$
             \cite{gocksch,MTc}. 
             The data have been rescaled by the appropriate ratio 
             of $2 \pi$, eq.~(\ref{freequarks}), to its finite
             lattice size corrected value so that the data can 
             immediately be compared with the continuum expectation 
             for free quarks shown as the horizontal line.
            }
  \end{center}
\end{figure}

Finally, in Figure~\ref{stagg} our results are compared with
data obtained in the staggered fermion discretization,
both below $T_c$ \cite{Boyd1} and above \cite{gocksch,MTc}.
Below $T_c$ the $\pi$ and $\rho$ screening masses 
are shown in the chiral limit. The agreement between the two
discretizations is evident. 
Above $T_c$ the staggered data
was computed at a small bare quark mass of $m_q a = 0.02$
while the Wilson results have been extrapolated to the chiral
limit. From Figure~\ref{mass12}, however, recall that above
$T_c$ the screening masses are practically independent of $m_q$
at small quark masses. Both data sets have been rescaled by
the appropriate ratio of $2 \pi$, eq.~(\ref{freequarks}), to its finite
lattice size corrected value so that the data can immediately be compared
with the continuum expectation for free quarks.
Figure~\ref{stagg} shows that above $T_c$ the $\rho$ screening mass
rapidly approaches $2 \pi$. This holds in both discretizations.
The pion however behaves very differently for the two lattice
actions. In the Wilson case, at $1.63 \, T_c$ the pion screening mass
is almost degenerate with the $\rho$ and close to $2 \pi$,
with a $\pi$ to $\rho$ ratio of 0.95(1).
The same approximate degeneracy of $\pi$ and $\rho$ screening masses
has been observed in the only other quenched analysis with Wilson quarks
available so far
\cite{QCDTARO}. Here a ratio of 0.955(7) has been obtained
at a temperature of about $1.5 \, T_c$.
Quite contrary,
the staggered pion screening mass is much smaller than the $\rho$,
$m_\pi / m_\rho = 0.75(2)$ at $1.8 \, T_c$,
and even at
$5 \, T_c$ reaches only about 75 \% of two times the lowest
Matsubara frequency. 
There is no immediate explanation at our hands at least.
Possible reasons include the different symmetries of the
lattice actions at non-vanishing lattice spacing
although the lattice spacings are below 0.1 fm,
the different aspect ratios although it is hard to see why this
should affect the $\pi$ to $\rho$ ratio, and finite volume
effects although those would be expected to go the opposite way.
Thus, it is hoped that future systematic studies help to resolve
this discrepancy between the two lattice fermion formulations.

\section{Dispersion relation}
\label{sec:disp}

At non-zero temperature, Lorentz invariance 
is broken because the temporal direction is
distinguished as the direction of the four-velocity
of the heat bath.
As a consequence, 
unlike the zero temperature case where it depends
on the Lorentz invariant scalar $p^2$,
in this case the spectral density will depend
on temporal and spatial components $p_0$ and
$\vec p$ separately.
At temperatures below the confinement-deconfinement transition
the spectrum will still consist of particle excitations, yet,
their dispersion relations might be more complicated 
and reflect the breaking of Lorentzian invariance.
The spectral density will be of the form
\beq
\rho(p_0,\vec p) = 2 \pi \epsilon (p_0) 
\delta(p_0^2 - \omega^2(\vec p ,T))
\eeq
where 
\beq
\omega^2(\vec p ,T) = m^2 + {\vec p}^{\, 2} + \Pi(\vec p ,T)
\eeq
contains the temperature dependent vacuum polarization tensor
$\Pi(\vec p ,T)$.
As a simple example, assume that the temperature effects can
be absorbed into a temperature dependent mass $m(T)$ and a
coefficient $A(T)$ which might also be temperature dependent
and different from 1,
\beq
\omega^2(\vec p ,T) \simeq m^2(T) + A^2(T) {\vec p}^{\, 2}
\eeq
Such an approximation might hold at least at small temperatures.  
In this case, at zero momentum the temporal correlator will decay 
with the so-called pole mass $m(T)$
\beq
C(\vec p = 0,t) \sim \exp(-m(T) t)
\eeq
whereas the spatial correlation function has an exponential
fall-off 
\beq
C(\vec {\tilde p} = 0,z) \sim \exp(-m_{\rm sc}(T) z)
\eeq
determined by the screening mass $m_{\rm sc}(T) = m(T)/A(T)$
which differs from the pole mass if $A(T) \neq 1$.
At non-vanishing ``momentum'', $\vec {\tilde p} \neq 0$, the
exponential decrease of the spatial correlator is described
by $\omega_{\rm sc}$,
\beq
C(\vec {\tilde p},z) \sim \exp(-\omega_{\rm sc} z)
\eeq
where in this particularly simple example $\omega_{\rm sc}$ is given as
\beq
\omega_{\rm sc}^2 = {\vec p}_\perp^{\,2} + \frac{\omega_n^2}{A^2}
                     + m_{\rm sc}^2
\label{Lorentz}
\eeq
Comparing different projections to Matsubara frequencies
$\omega_n = 2 \pi Tn$ and to spatial momenta ${\vec p}_\perp$
might thus reveal
a dispersion relation different from the zero temperature one
and, moreover, indicate a coefficient $A$ different from 1
and correspondingly a difference between pole and screening mass.

The formulae given above are the continuum dispersion relations.
They are modified on the lattice. Zero temperature studies
\cite{Rajan,rom1157} have shown that a lattice dispersion of
the form
\beq
\sinh^{2}\left(\frac{E}{2}\right) = 
\sum_{k} \sin^{2}\left(\frac{p_{k}}{2}\right) 
+ \sinh^{2}\left(\frac{M}{2}\right)
\label{latticedisp}
\eeq
arising from an effective boson action with a nearest-neighbor
kinetic term, is best capable to describe lattice data at 
non-zero momentum.

\begin{figure}[h]
  \begin{center}
    \epsfig{file=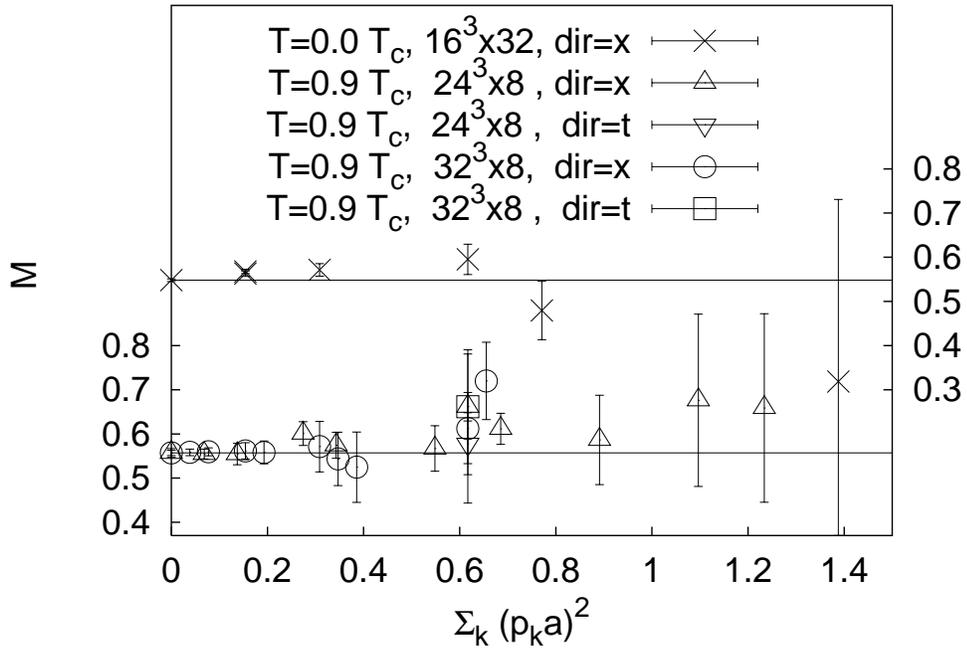,width=134mm}
    \caption{The lattice dispersion relation shown for the
             pseudoscalar channel at $\kappa = 0.141$. In the figure,
             ``{\tt dir}'' indicates whether spatial or temporal
             components of $(p_x,p_y,p_t)$ were chosen
             to be different from 0. The horizontal lines
             denote the dispersion relation eq.~(\ref{latticedisp}).
             Note that the $T=0$ data has been shifted upwards,
             right scale.
    \label{dispersion}
    }
  \end{center}
\end{figure}

We investigated the lattice dispersion relation of mesons below $T_c$, 
$T \simeq 0.93 \, T_c$,
for different quark masses and compared it with $T = 0$ data.
In Figure~\ref{dispersion} we show our results obtained 
in the pseudoscalar channel at $\kappa = 0.141$.
We have plotted $M_\pi a$
obtained by subtracting the momentum contribution
$\sin^2(p_ka/2)$ from both sides of eq.~(\ref{latticedisp})
where $E$ is the fitted coefficient of the exponential fall-off
in $z$ of the spatial correlation function.
Results at other $\kappa$ values investigated are very similar.

In Figure~\ref{dispersion}, a $p_k$ independence of the data
indicates that eq.~(\ref{latticedisp}) is the correct
dispersion relation. Indeed, it seems to be favored by the
zero temperature data. As in \cite{Rajan}, using a different
dispersion
relation led to data points rising or falling with $p_k$.

Regarding the data at $T = 0.93 \, T_c$, again we observe
that the zero-temperature dispersion, eq.~(\ref{latticedisp}),
is describing the data for non-vanishing spatial momentum
components properly. Note that the figure contains data
from two different lattice sizes, leading to different values
for the spatial momenta.
Moreover, the single data point for
the lowest bosonic Matsubara frequency $\omega_1 = 2 \pi T$
is also lying on the horizontal line, suggesting that
the coefficient $A$ of eq.~(\ref{Lorentz}) is not too far
from unity.
Of course, this conclusion is tied to the applicability of
eq.~(\ref{Lorentz}), moreover, the statistical significance is
certainly not overwhelming.
Nevertheless, this result might be taken as 
further support for 
the difference between pole and screening masses 
not being too large at $T \simeq 0.93 \, T_c$,
see also \cite{QCDTARO}.

\begin{figure}[t]
  \begin{center}

\vspace*{-5mm} 
    \epsfig{file=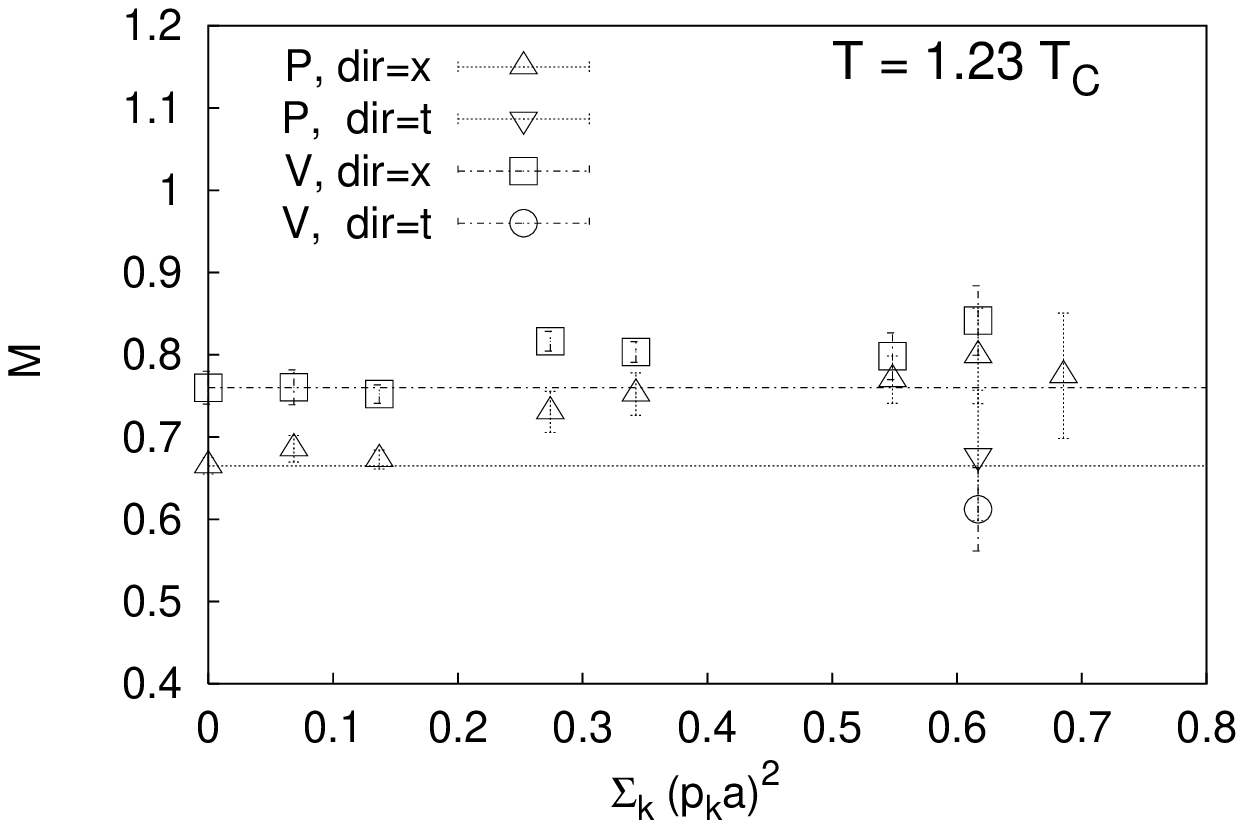,width=90mm}

\vspace*{-4mm}
    \epsfig{file=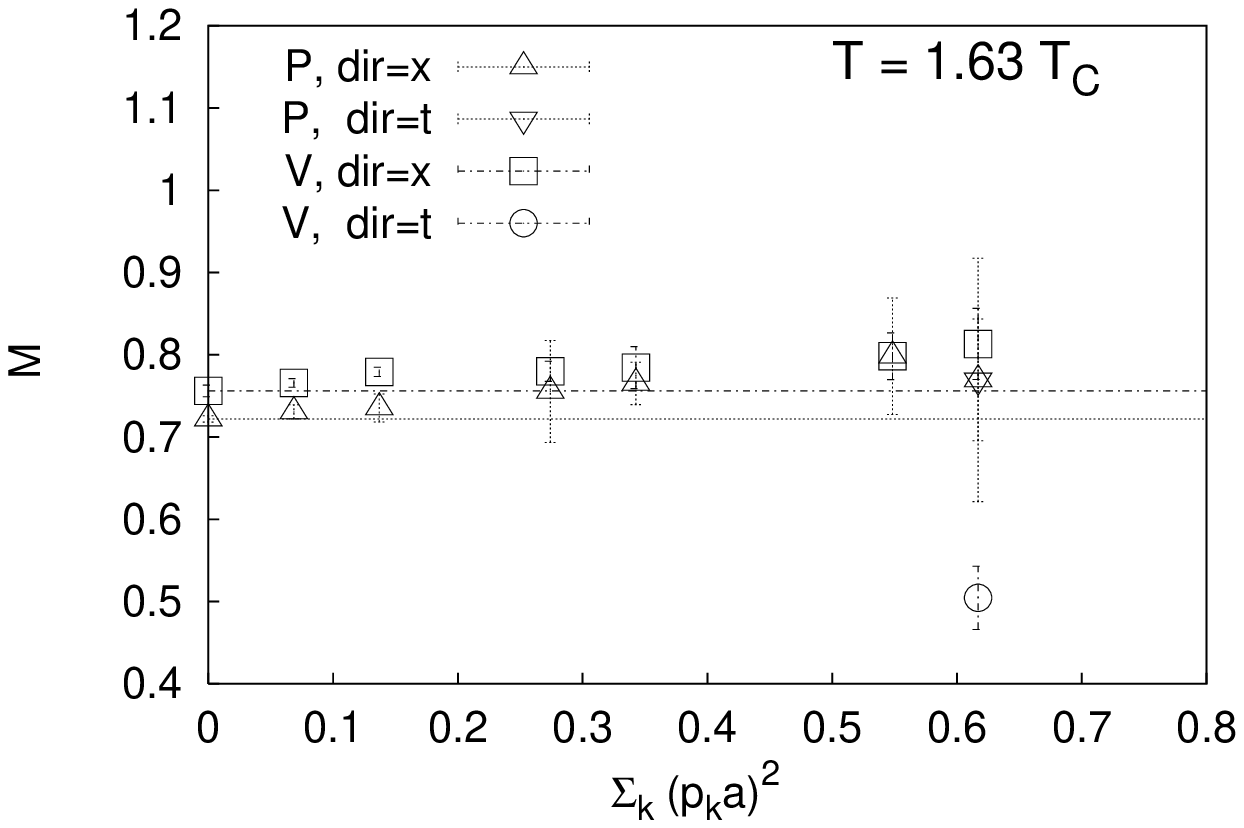,width=90mm}

\vspace*{-4mm}
    \epsfig{file=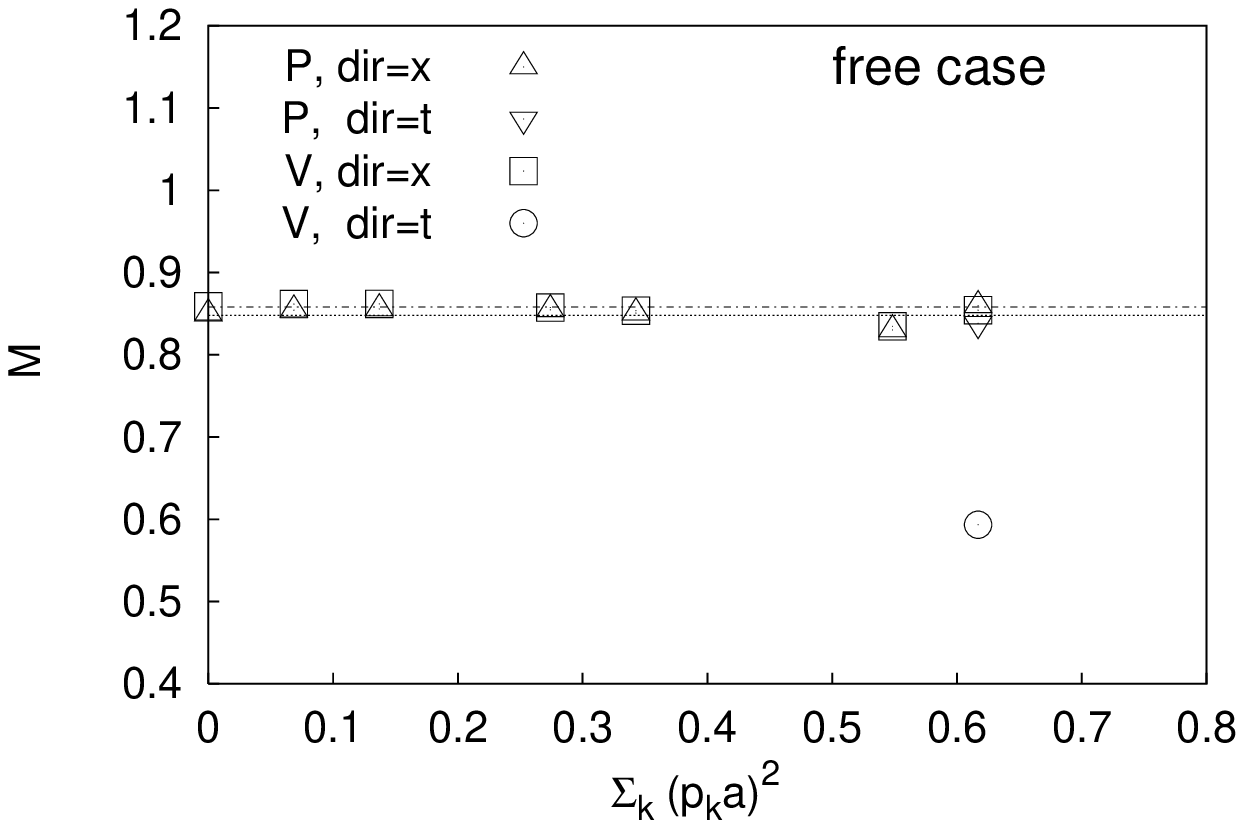,width=90mm}

\vspace*{-2mm}
    \caption{
\label{disp_high_t}
             The lattice dispersion relation at two temperatures
             above $T_c$ and in the free quark case.
             Shown are both the results for the pseudoscalar and
             the vector channel. The bare quark mass is roughly equal
             to $m_s$.
             Again,
             ``{\tt dir}'' indicates whether spatial or temporal
             components of $(p_x,p_y,p_t)$ were chosen
             to be different from 0. The horizontal lines
             denote the dispersion relation eq.~(\ref{latticedisp}).
             }
  \end{center}
\end{figure}

In addition to the investigation below $T_c$, we also computed
the dispersion relation in mesonic channels in the deconfined
phase. The results are shown and compared with the
free quark case in Figure \ref{disp_high_t}.
The former have been determined by fits in the same way as
the masses while
for the latter, to be definite we use the effective energies at $N_\sigma/4$,
obtained from sink operators with $R=4$. Both, Monte Carlo
as well as free quark results 
are somewhat dependent on the quark-antiquark distance.
As can be read off the figure, at $1.23 \, T_c$ the data
for spatial momenta
show systematic deviations from the preferred
zero-temperature dispersion relation. These differences have almost
disappeared at $1.63 \, T_c$ and the data are becoming consistent
with the zero-temperature dispersion relation which also happens
to describe the free quark case.
For the lowest temporal mesonic ``momentum'', $p_t = 2 \pi T$,
at $1.23 \, T_c$ pseudoscalar and vector ``energy'' both are
systematically lower than the corresponding results for a spatial
momentum of exactly the same value showing that Lorentzian
symmetry is disturbed. At $1.63 \, T_c$ the data follow the
free quark behavior, in particular insofar in the vector channel
the ``energy'' at $p_t = 2 \pi T$
is much smaller than the ``energy'' at
the same spatial momentum\footnote{Note that this, at first sight
strange behavior can easily be reproduced in the continuum if only
the lowest quark Matsubara frequency would contribute to the
correlation function. Of course, in this case pseudoscalar and vector
channel would lead to the same result. Apparently, the interplay
between higher quark momenta and the spin structure is important.}.

\section{Wave function}
\label{sec:wavefunc}

Bethe-Salpeter amplitudes provide information about the probability
of finding a pre-arranged configuration of quarks
inside a hadron. Their general definition is given by
\beq
\Phi(\vec R) = \langle 0 | {\cal O}(\vec R) | H(\vec p) \rangle
\eeq
where, for $| H \rangle$ being a meson state, the operator  
${\cal O}(\vec R)$ annihilates a quark-antiquark pair with the
appropriate quantum numbers separated by $\vec R$.
The choice of the operator is not unique.
On the lattice, Bethe-Salpeter amplitudes have been studied
in Coulomb and in Landau gauge 
or by using various gauge-invariant definitions
\cite{Velikson,Chu,Hecht,Rajan_BS,Lacock}.
The various methods treat the gluon flux tube connecting quark
and antiquark in a different way so that it may not be surprising
that different definitions of the amplitudes have lead to different
results (see however \cite{Rajan_BS}). Moreover, an immediate
connection between the size of a hadron as extracted from
the amplitudes and the measured radii has not been demonstrated.
Nevertheless, the Bethe-Salpeter amplitudes do provide some
qualitative insight into the quark distribution inside a hadron,
and in particular they should be able to capture differences
in the structure of hadronic excitations at zero and high
temperature if these are present \cite{MILC_BS}.

As a by-product of the attempt to obtain a better projection
onto the ground state, in this paper we use a gauge-invariant
definition of ${\cal O}(\vec R)$ as given in eq.~(\ref{operator})
\cite{Lacock}. We analyze S-wave states only as this operator
is invariant under $90^\circ$ rotations.
The Bethe-Salpeter amplitudes are determined from simultaneous
two-state fits to all the correlation functions $C_R$ at all $R$ values
at which they were calculated. In these fits the ground state
mass was required to be the same at all $R$ whereas the mass of
the second state as well as the amplitudes for both ground and
excited state were allowed to be $R$ dependent.
Again, the fit interval was varied and we checked for stability
of the fit results under this variation as well as for agreement
of the ground state mass with the numbers obtained as explained
in section~\ref{sec:masses}.

\begin{figure}[h]
  \begin{center}
    \epsfig{file=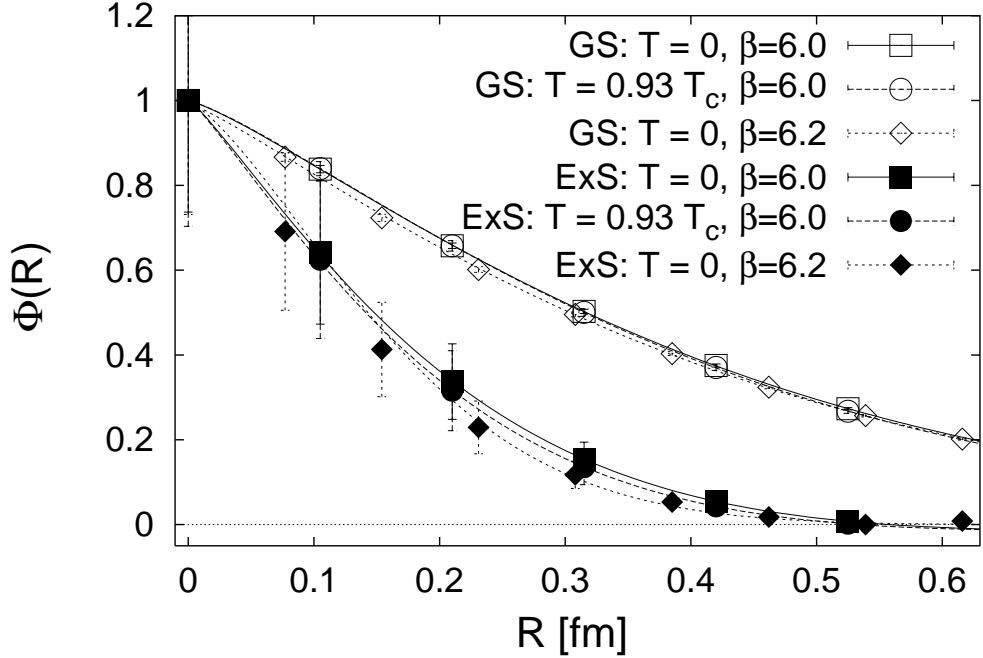,width=134mm}
    \hspace*{-1mm}
    \caption{\label{wavefunction}
    Wave functions for the ground (GS) and excited state (ExS) of the
    pseudoscalar at $T \! = \! 0$ and $T \! = \! 0.93  T_c$, 
    for $m_{q} \! \approx \! m_{s}$ }
  \end{center}
\end{figure}

The procedure proved stable enough to extract the Bethe-Salpeter
amplitude for the ground state as well as to obtain an estimate
for the excited state.
An example of the results 
is shown in Figure~\ref{wavefunction}.
Here, the wave functions have been normalized to 1 at $R=0$.
Note that the amplitude of the excited state is vanishing
at about 1/2 fm.
This corresponds with the observation that 
at large $R$
effective masses approach the plateaux in effective 
mass plots from below, see Figure~\ref{projection}.
The results for the excited states are more sensitive
to variations of the fit interval since at small separations from the source
more than one excited state can contribute whereas at large
separations the excited state dies out. We will thus concentrate
on the ground state results in the following.

The ground state wave functions \footnote{The excited states
could be well fitted by the ansatz
$
\Phi(R) \propto ( 1 - R/b_1) \exp(-(R/a_1)^\nu)
$
which accounts for the node in the wave function.
}
are fitted with slightly modified hydrogen
S-wave functions \cite{Hecht},
\beq
\Phi(R) ~ \propto ~ \exp(-(R/a_0)^{\nu})
\label{gs_wave}
\eeq
with the free parameters $a_0$ and $\nu$. 
In non-relativistic potential models the power of $R$ is obtained
as $\nu=1$ for a purely Coulombic and $\nu = 3/2$ for
a linear potential.
The wave functions can be used to compute an estimate of the
size of the meson \cite{Hecht},
\beq
\langle r^2 \rangle = \int d^{\,3}\vec R \left(\frac{R}{2}\right)^2 \Phi^2(R)
                    / \int d^{\,3}\vec R \;\, \Phi^2(R)
               = \frac{a_0^2 \, \Gamma(5/\nu)}{4^{1+1/\nu} \,\Gamma(3/\nu)}.
\label{rms}
\eeq

\begin{table}[t]
\begin{center}
\begin{tabular}{|@{}c@{\hspace*{2mm}}c@{\hspace*{2mm}}l||l@{\hspace*{1mm}}|l|l||l|l|l|}
\hline
\multicolumn{3}{|c||}{\rule[-3mm]{0mm}{8mm} } &
\multicolumn{3}{|c||}{\textbf P} &
\multicolumn{3}{|c|}{\textbf V} \\
\cline{4-9}
\rule[-3mm]{0mm}{8mm}
size & $\beta$ & \multicolumn{1}{c||}{$\kappa$} & 
                 \multicolumn{1}{|c|}{$a_0$} & 
                 \multicolumn{1}{|c|}{$\nu$} & 
                 \multicolumn{1}{|c||}{$\langle r^2 \rangle$} &
                 \multicolumn{1}{|c|}{$a_0$} & 
                 \multicolumn{1}{|c|}{$\nu$} & 
                 \multicolumn{1}{|c|}{$\langle r^2 \rangle$} \\
\hline
\hline
\rule[-0mm]{0mm}{5mm}
$16^3\times 32$ & 6.0 & 0.141 & 4.06(4) & 1.24(4) & 6.7(5) & 5.35(4) & 1.42(4) & 8.7(5) \\
                &     & 0.130 & 3.42(2) & 1.29(4) & 4.3(3) & 4.06(2) & 1.42(5) & 5.0(3) \\
                &     & 0.128 & 3.38(4) & 1.23(4) & 4.7(4) & 4.03(5) & 1.32(6) & 5.7(6) \\
\hline
\rule[-0mm]{0mm}{5mm}
$24^3\times 48$ & 6.2 & 0.141 & 5.32(3) & 1.15(2) &13.9(7) & 7.59(5) & 1.40(3) &18.0(8) \\
                &     & 0.130 & 4.27(3) & 1.20(3) & 8.0(5) & 5.27(5) & 1.34(5) & 9.5(8) \\
                &     & 0.128 & 4.09(3) & 1.20(3) & 7.4(5) & 4.97(4) & 1.34(4) & 8.4(5) \\
\hline
\hline
\rule[-0mm]{0mm}{5mm}
$32^3\times 8$  & 6.0 & 0.145 & 4.17(7) & 1.21(6) & 7.5(10)& 5.80(14) & 1.50(14) & 9.4(13) \\
                &     & 0.1445& 4.15(6) & 1.21(4) & 7.4(6) & 5.87(11) & 1.51(22) & 9.3(24) \\
                &     & 0.144 & 4.15(6) & 1.23(6) & 7.1(9) & 5.64(13) & 1.47(17) & 9.1(19) \\
                &     & 0.141 & 4.03(3) & 1.25(4) & 6.5(5) & 5.36(5)  & 1.50(14) & 7.9(13) \\
\hline
\rule[-0mm]{0mm}{5mm}
$24^3\times 8$  & 6.2 & 0.1428  & 3.90(17) & 1.38(23) & 4.9(17) & 4.85(9)  & 1.81(32) & 4.9(11) \\
                &     & 0.14232 & 3.94(8)  & 1.42(14) & 4.7(9)  & 4.74(18) & 1.76(30) & 4.9(11) \\
                &     & 0.14151 & 3.84(7)  & 1.43(11) & 4.4(5)  & 4.75(8)  & 1.76(30) & 4.9(10) \\
                &     & 0.136   & 3.70(5)  & 1.46(10) & 3.9(5)  & 4.39(8)  & 1.68(22) & 4.5(8)  \\
                &     & 0.128   & 3.47(3)  & 1.55(12) & 3.1(4)  & 3.90(3)  & 1.70(13) & 3.5(4)  \\
\hline
\rule[-0mm]{0mm}{5mm}
$24^2 \, 64 \times 8$  & 6.4 & 0.1409  & 3.91(5)  & 1.59(8)  & 3.8(3)  & 4.86(3)  & 1.89(19) & 4.7(5)  \\
                &     & 0.1406  & 3.92(5)  & 1.60(7)  & 3.8(3)  & 4.86(4)  & 1.88(20) & 4.7(6)  \\
                &     & 0.1403  & 3.92(5)  & 1.60(7)  & 3.8(3)  & 4.86(3)  & 1.88(20) & 4.7(6)  \\
                &     & 0.14    & 3.93(6)  & 1.61(9)  & 3.8(3)  & 4.85(2)  & 1.81(19) & 4.9(6)  \\
                &     & 0.13    & 3.69(3)  & 1.69(7)  & 3.1(2)  & 4.19(1)  & 1.82(9)  & 3.7(2)  \\
\hline
\end{tabular}
\vspace*{-1.2mm}
\caption{\label{tab:radius}
         Results for the coefficients $a$ and $\nu$ of the ground state wave
         function, eq.~(\ref{gs_wave}), as well as for the mean radius squared,
         eq.~(\ref{rms}), for the pseudoscalar (P) and vector meson (V).}
\end{center}
\end{table}

The results of these fits to the pseudoscalar as well as
vector meson ground state are summarized in Table~\ref{tab:radius}.
We first want to stress that the measured zero temperature wave functions,
i.e. the parameter $a_0$ and the radius respectively,
scale with $\beta$ as the lattice spacing
when the (physical) quark mass is fixed. 
This is lending further support that the investigated 
observable is physical. 

At $T = 0$,
the obtained values for the parameter $\nu$ lie between
the values for a purely Coulombic and a purely linear potential.
The values for $\nu$ are slightly increasing with the
quark mass and they are larger for the vector meson than
for the pseudoscalar.
Moreover, the values for $a_0$ and for the radii are also increasing
with the quark mass and again are larger for the vector meson
than for the pseudoscalar. Both effects are consistent
with the experimental observation that the $\rho$ meson is
larger than the pion and thus probes the linear rising
part of the potential at larger distances.
Compared to the (Coulomb gauge) results of \cite{Hecht} at the same
pseudoscalar to vector meson mass ratio, we observe
perfect agreement for the parameter $\nu$ of the 
pseudoscalar ground state
whereas our $\nu$ value for the vector meson is slightly smaller.
The second moments of the meson wave functions,
eq.~(\ref{rms}),
are about 10 \% larger than in \cite{Hecht}. Still,
as reported there, we obtain radii which are 
about a factor two smaller than sizes obtained
from form factor measurements.
Whether this discrepancy is due to the quenched
approximation i.e. to not taking proper account of a
pion cloud still has to be speculated at the present stage.

At $T=0.93 T_c$ the pseudoscalar as well as vector meson
(spatial) wave functions agree with the 
(temporal) zero temperature ones at quark masses where
we have data for both. This
also holds for the excited state, see Figure~\ref{wavefunction}.
At this temperature we also have data at small quark masses
so that we can attempt a chiral extrapolation. A fit to the
squared radii which is linear in $m_{PS}^2$ leads to values
of $\langle (r/a)_\pi^2 \rangle = 7.6(6)$ and
$\langle (r/a)_\rho^2 \rangle = 9.6(12)$, respectively,
which are in agreement within errors with the 
zero-temperature numbers reported
in \cite{Hecht}.

\begin{figure}[h]
  \begin{center}
    \epsfig{file=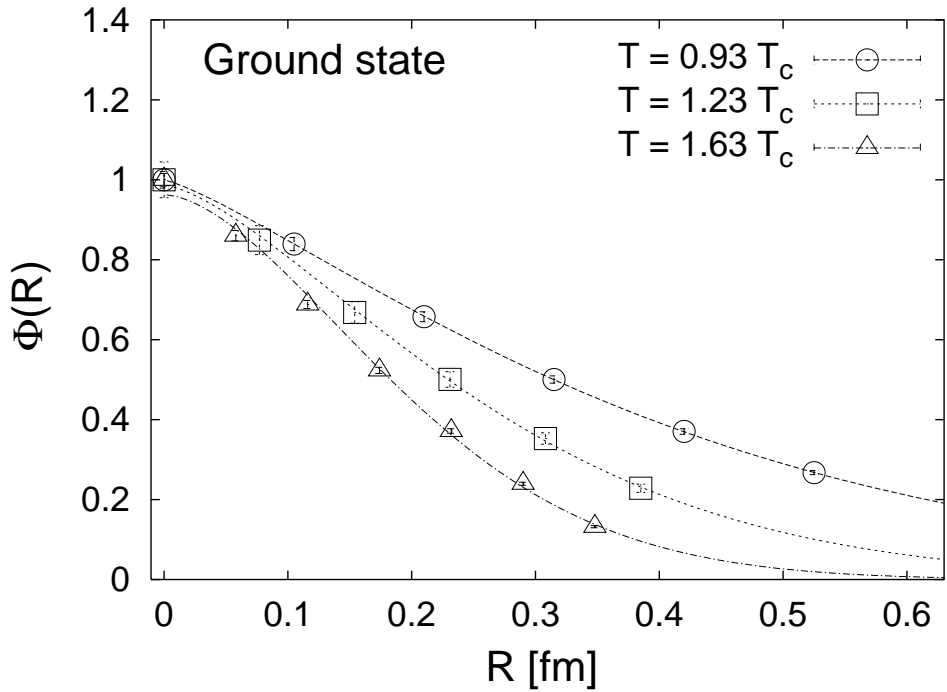,width=134mm}
    \hspace*{-1mm}
    \caption{\label{wavefunction_t}
    Wave functions for the ground state of the
    pseudoscalar at $T \! \neq \! 0$, 
    for $m_{q} \! \approx \! m_{s}$}
  \end{center}
\end{figure}

At the two temperatures above $T_c$ it is observed that the spatial
wave functions in lattice units practically remain unchanged
between the two $\beta$ values.
This is distinctively different from the zero temperature
situation.

In more detail, fits to the wave functions with eq.~(\ref{gs_wave})
return the values given in Table~\ref{tab:radius}.
The results for the parameter $a_0$ are almost independent
of $\beta$. The parameter $\nu$ takes on values which are larger
than at zero temperature. Close to the chiral limit, $\nu$
changes from about 1.4 at $1.23 T_c$ to 1.6 at $1.63 T_c$ for the
pion while for the $\rho$ the change with temperature is smaller,
$\nu \simeq 1.8$ and 1.9 respectively.
The second moment of the wave function,
$\langle r^2 \rangle$, in lattice units
stays constant within errors for the $\rho$
while for the pion with rising temperature we observe a
decrease of the central values, with
error bars just overlapping.
As the quark mass is increased the size of the state
is slightly shrinking.

Translated to physical distances 
a shrinking of the states' size with temperature
is clearly oberved,
Figure~\ref{wavefunction_t}.
At least qualitatively, this observation is
in accord with
the anticipated behavior \cite{Hansson,Koch} of
\beq
\Phi(R) \propto \exp( - \sqrt{\sigma_{\rm spat}(T) \pi T} R^{3/2})
\eeq
where $\sigma_{\rm spat}(T)$ is the temperature dependent
string tension of spatial Wilson loops,
$\sqrt{\sigma_{\rm spat}(T)} = c \, g^2(T)T$.
Given the small variation of the effective coupling constant
$g(T)$ between $T=1.23$ and $1.63 T_c$, 
$\langle r^2 \rangle$ should thus shrink proportional to $1/T^2$ to
first approximation. Indeed, this seems to hold for the
vector meson whereas the variation with temperature of
the pion radius is stronger. Moreover, the power of the $R$
dependence is not quite as anticipated. A more detailed
investigation along the lines of \cite{Koch} is planned for the
future.

\section{Conclusion}

In this paper we have presented results of a lattice investigation
of meson correlation functions built from improved Wilson quarks
on quenched gauge field configurations. Below the confinement-deconfinement 
transition temperature we do not observe any significant difference
between finite temperature screening masses and zero temperature
masses. This agrees with earlier findings obtained in the
staggered discretization.
Likewise, the spatial wavefunctions at non-vanishing temperature
agree with wave functions at $T = 0$.
Moreover, the dispersion relation does not seem to depend on
whether the ``momenta'' are chosen to point in the spatial or
temporal direction. This can be viewed as a hint towards the
similarity of screening and pole masses as has also been reported
in an earlier study on anisotropic lattices.
Above the critical temperature, we have computed
meson screening masses which in both cases,
pseudoscalar as well as vector channel, are close to twice
the lowest quark Matsubara frequency.
Moreover, at $1.63 \, T_c$ the ``momentum'' dependencies of
the correlation functions are in agreement with the free quark case
while at $1.23 \, T_c$ the situation certainly is more complex.
Finally, the spatial wave functions, at least qualitatively, 
behave in accord
with the expectation that the temperature dependence of the
spatial string tension leads to a narrowing of the wave functions
with rising temperature. We thus come to conclude that above $T_c$
the spatial correlation functions are consistent with a leading
two quark contribution. This still leaves room for infrared
non-perturbative effects. Note in this respect that HTL-resummed
perturbation theory does not show significant changes in (temporal)
correlation functions although the underlying spectral function
is quite different from the free case \cite{HTL}. It is our hope
that more refined analysis techniques along the lines set in
\cite{MEM} and perhaps more systematic studies of infrared
properties on large lattices extend the knowledge about
hadronic excitations in the plasma phase.

\section*{Acknowledgement}
This work was supported by the TMR network ERBFMRX-CT-970122 and
the DFG grant Ka 1198/4-1.
The numerical work has been carried out
on the CRAY T3E at the Konrad Zuse Institut Berlin
and to a small part on the CRAY T3E at NIC, FZ J\"ulich.
We thank both institutions for allocation of computer time
and
H. St\"uben as well as N. Attig for their generous help.


\begin{thebibliography}{99}
\bibitem{dilepton_exp} G. Akakichiev et al. (CERES Collaboration),
        Phys. Lett. B422 (1998) 405.
\bibitem{Rapp} R. Rapp and J. Wambach, {\it ``Chiral Symmetry 
        Restoration and Dileptons in Relativistic Heavy-Ion Collisions''},
        hep-ph/9909229.
\bibitem{DeTar} C. DeTar, \PR D32 (1985) 276; \PR D37 (1987) 2328.
\bibitem{Detar2} C. DeTar and J. B. Kogut, Phys. Rev. Lett. 59 (1987) 399.
\bibitem{gocksch} A. Gocksch, P. Rossi and U.M. Heller, \PL B205 (88) 334.
\bibitem{Gottlieb} S. Gottlieb et al., Phys. Rev. Lett. 59 (1987) 1881.
\bibitem{MTc} K. D. Born et al., Phys. Rev. Lett. 67 (1991) 302.
\bibitem{Sourendu} S. Gupta, Phys. Lett. B288 (1991) 171.
\bibitem{Boyd3} G. Boyd, S. Gupta and F. Karsch, Nucl. Phys. B385 (1992) 482.
\bibitem{Bernard2} C. Bernard et al., Phys. Rev. D45 (1992) 3854.
\bibitem{Boyd2} G. Boyd et al., Z. Phys. C64 (1994) 331.
\bibitem{Boyd1} G. Boyd et al., Phys. Lett. B349 (1995) 170.
\bibitem{Gottlieb2} S. Gottlieb et al., 
              Phys. Rev. D47 (1993) 3619,
              Phys. Rev. D55 (1997) 6852.
\bibitem{Lagae} J.B. Kogut, J.-F. Laga\"e and D.K. Sinclair,
              Phys. Rev. D58 (1998) 54504.
\bibitem{Nucu} T. Hashimoto, T. Nakamura and I. O. Stamatescu,
              Nucl. Phys. B400 (1993) 267.
\bibitem{Nucu2} T. Hashimoto, T. Nakamura and I. O. Stamatescu,
              Nucl. Phys. B406 (1993) 325.
\bibitem{QCDTARO} Ph. de Forcrand et al. (QCD-TARO),
              Phys. Rev. D63 (2001) 054501.
\bibitem{Columbia} P. Vranas (QCDSP Coll.), 
              Nucl. Phys. Proc. Suppl. 83 (2000) 414. 
\bibitem{MILC_BS} C. Bernard et al., \PRL 68 (1992) 2125.
\bibitem{Koch} V. Koch et al., Phys. Rev. D46 (1992) 3169.
\bibitem{Borgs} C. Borgs, Nucl. Phys. B261 (1985) 455.
\bibitem{Janos} E. Manousakis and J. Polonyi,
                   Phys. Rev. Lett. 58 (1987) 847.
\bibitem{Martin} F. Karsch, E. Laermann and M. L\"utgemeier,
                   Phys. Lett. B346 (1995) 94.
\bibitem{Hansson} T.H. Hansson and I. Zahed, Nucl. Phys. B374 (1992)
                277.
\bibitem{sheik} B. Sheikholeslami and R. Wohlert, \NP B259 (1985) 572.
\bibitem{CM} N. Cabibbo and E. Marinari, Phys. Lett. 119B (1982) 387.
\bibitem{FHKP} K. Fabricius and O. Haan, Phys. Lett. 143B (1984) 459;\\
       A. Kennedy and B. Pendleton, Phys. Lett. 156B (1985) 393.
\bibitem{Adler} S. Adler, Phys. Rev. D23 (1981) 2901.
\bibitem{Klassen} R.G. Edwards, U.M. Heller and T.R. Klassen,
              Nucl. Phys. B517 (1998) 377 and references therein.
\bibitem{Beinlich} B. Beinlich et al., Eur. Phys. J. C6 (1999) 133.
\bibitem{Paige} C.C. Paige and M.A. Saunders, \JNA 12 (1975) 617.
\bibitem{Frommer} A. Frommer et al., Int. J. Mod. Phys. C5 (1994) 1073
              and references therein.
\bibitem{edin526} UKQCD Coll., \PR D49 (1994) 1594.
\bibitem{rom1157} C. Allton et al., \NP B489 (1997) 427.
\bibitem{edin524} UKQCD Coll., \PR D49 (1994) 474.
\bibitem{Lacock} UKQCD Coll., \PR D51 (1995) 6403.
\bibitem{APEsmear} M. Albanese et al., \PL 192B (1987) 163.
\bibitem{Rajan} T. Bhattacharya et al., \PR D53 (1996) 6486.
\bibitem{APEsmall} C. Allton et al. and APE Coll., \NP B413 (1994) 461.
\bibitem{Velikson} B. Velikson and D. Weingarten, \NP B249 (1985) 433.
\bibitem{Chu} M.-C. Chu, M. Lissia and J. Negele, \NP B360 (1991) 31.
\bibitem{Hecht} M.W. Hecht and T.A. DeGrand, \PR D46 (1992) 2155
\bibitem{Rajan_BS} R. Gupta, D. Daniel and J. Grandy, \PR D48 (1993) 3330.
\bibitem{HTL} F. Karsch, M.G. Mustafa and M.H. Thoma,
              Phys. Lett. B497 (2001) 249.
\bibitem{MEM} M. Asakawa, T. Hatsuda and Y. Nakahara, 
              hep-lat/0011040; \\
              I. Wetzorke and F. Karsch, hep-lat/0008008.

\end{thebibliography}
\end{document}